\documentclass[twocolumn, showpacs,preprintnumbers,amsmath,amssymb]{revtex4}
\usepackage{amssymb}
\usepackage{xcolor}
\usepackage{graphicx}
\usepackage{dcolumn}
\usepackage{bm}
\usepackage{multirow}
\usepackage{booktabs}
\usepackage{CJK}
\usepackage{natbib}
\usepackage{soul}
\usepackage{amssymb}

\begin{document}
\preprint{APS/123-QED}

\title{Engineering long-range interactions between ultracold atoms with light}
\begin{CJK*}{GBK}{song}\end{CJK*}

\author{\begin{CJK*}{GBK}{song}T. Xie$^{1,2}$\end{CJK*}}%
\author{\begin{CJK*}{GBK}{song}A. Orb\'{a}n$^3$\end{CJK*}}
\author{\begin{CJK*}{GBK}{song}X. Xing$^1$\end{CJK*}}
\author{\begin{CJK*}{GBK}{song}E. Luc-Koenig$^1$\end{CJK*}}
\author{\begin{CJK*}{GBK}{song}R. Vexiau$^1$\end{CJK*}}
\author{\begin{CJK*}{GBK}{song}O. Dulieu$^1$\end{CJK*}}
\author{\begin{CJK*}{GBK}{song}N. Bouloufa-Maafa$^1$\end{CJK*}}
\affiliation{$^1$Universit\'e Paris-Saclay, CNRS, Laboratoire Aim\'{e} Cotton, 91405 Orsay, France}%
\affiliation{$^2$State Key Laboratory of Molecular Reaction Dynamics, Dalian Institute of Chemical Physics,
Chinese Academy of Sciences, Dalian, Liaoning 116023, China}%
\affiliation{$^3$Institute for Nuclear Research(ATOMKI), H-4001 Debrecen, Pf. 51, Hungary}%
\date{\today}

\begin{abstract}
Ultracold temperatures in dilute quantum gases opened the way to an exquisite control of matter at the quantum level. Here we focus on the control of ultracold atomic collisions using a laser to engineer their interactions at large interatomic distances. We show that the entrance channel of two colliding ultracold atoms can be coupled to a repulsive collisional channel by the laser light so that the overall interaction between the two atoms becomes repulsive: this prevents them to come close together and to undergo inelastic processes, thus protecting the atomic gases from unwanted losses. We illustrate such an optical shielding mechanism with $^{39}$K and $^{133}$Cs atoms colliding at ultracold temperature (\textless 1 microkelvin). The process is described in the framework of the dressed-state picture and we then solve the resulting stationary coupled Schr\"{o}dinger equations. The role of spontaneous emission and photoinduced inelastic scattering is also investigated as possible limitations of the shielding efficiency. We predict an almost complete suppression of inelastic collisions over a broad range of Rabi frequencies and detunings from the $^{39}$K D2 line of the optical shielding laser, both  within the [0, 200~MHz] interval. We found that the polarization of the shielding laser has a minor influence on this efficiency. This proposal could easily be formulated for other bialkali-metal pairs as their long-range interaction are all very similar to each other.
\end{abstract}


\maketitle

\section{Introduction}

Research focusing on the formation of ultracold molecular quantum gases is a continuously expanding field due to its envisioned applications such as quantum-controlled chemistry, or as a promising platform for quantum information and quantum technologies \cite{baranov2008,bloch2008,lahaye2009,bloch2012}. Among the wealth of molecular species, ultracold polar diatomic molecules composed of a pair of different alkali-metal atoms attract a lot of interest. First, alkali-metal atoms are among the most appropriate species for laser cooling and evaporative cooling, eventually leading to quantum degeneracy of the atomic gases: the most advanced investigations on ultracold molecules indeed concern species created by associating atoms from these ultracold atomic gases. Second, such heteronuclear molecules in their ground state exhibit a significant permanent electric dipole moment in their own frame,  which can generate considerable anisotropic dipole-dipole interactions in ultracold conditions \cite{doyle2004,carr2009b,dulieu2009,quemener2012,moses2017,bohn2017}. 

One of the main objectives of the researchers is the formation of ultracold polar molecules in their absolute ground state, namely when all degrees of freedom (electronic, vibrational, rotational, hyperfine) are set to their lowest state so that the molecular internal energy is minimal. Among the various species, a few of them are reactive in such a state (\textit{e.g.} KRb+KRb $\rightarrow$ K$_2$ + Rb$_2$ with an exothermiticity of about 10~cm$^{-1}$, and all the pairs containing Li atoms), and some others are expected to undergo only elastic collisions (\textit{e.g.} NaK, NaRb, KCs, RbCs). But the hopes for achieving a molecular degenerate quantum gas with the latter species face for several years an unexpected drawback in ongoing experiments: the number of ultracold molecules created in their absolute ground state is observed to decrease within milliseconds, despite that they are not expected to react with each other \cite{ospelkaus2010a,ni2010,takekoshi2014,molony2014,guo2016,gregory2019}. The exact reasons for such losses are still under investigation. It has been proposed that a sticking mechanism generated by the presumably huge density of states of the four-atom complex creates a long-lived complex \cite{mayle2013,croft2014}, which could then decay by other mechanisms like inelastic collisions or photoassisted processes induced by the trapping laser. However this hypothesis has been reconsidered \cite{christianen2019a,christianen2019b,jachymski2021} while recent experimental results yield to contradictory interpretations in this matter \cite{gregory2019,gregory2020,liu2020,bause2021,gersema2021}.

 Therefore, theorists came soon with the idea of engineering the long-range interaction between ultracold molecules using external fields. Various approaches have been proposed, based on static electric field \cite{avdeenkov2006a,quemener2010b,wang2015,quemener2016,gonzalez-martinez2017}, or microwave (mw) field \cite{karman2018,lassabliere2018,karman2020}, which have been beautifully demonstrated experimentally \cite{matsuda2020,li2021,anderegg2021}. In both cases, the shielding of short-range interactions results from the mixture of the rotational levels of the incoming ground state molecules. 
 
 A third option has been proposed by us using an optical field \cite{xie2020}, namely, which mixes the molecular ground state to an electronically-excited molecular state, typically occurring in the infrared or optical frequency domain. This so-called optical shielding  (OS) technique is the purpose of the present paper, devoted to OS of short-range interactions between alkali-metal atoms of different species.
 
\begin{figure}[!t]\centering
\resizebox{0.45\textwidth}{!}
{\includegraphics{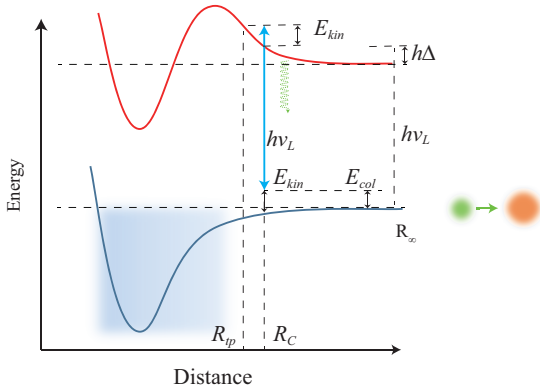}} 
\caption{The principle of the optical shielding (OS) in a classical view. Two ultracold ground-state atoms approach each other with a collision energy $E_{col}$. At the Condon point $R_C$, (\textit{i.e.} where the kinetic energy $E_{kin}$ is identical in the ground and excited states), the atom pair absorbs a photon from a laser with a frequency $\nu_L$ blue-detuned by $\Delta$ from an atomic transition, in order to reach a repulsive potential energy curve (PEC). The system evolves towards smaller distances until the turning point $R_{tp}$ (\textit{i.e.} where the kinetic energy $E_{kin}$ vanishes in the excited state). Back in $R_C$, the pair undergoes a stimulated emission towards the initial PEC, providing that the laser is powerful enough. The two atoms then drift away from each other, so that they never experience short-range interactions (shadowed area) where inelastic collisions could occur. Spontaneous emission from the upper PEC (green oscillating line) could limit the OS efficiency. Note that on this scheme, $E_{col}$, $E_{kin}$, $h\Delta$ are not on scale with respect to the PEC, for clarity.}
\label{fig:OSscheme}
\end{figure}

 OS was firstly applied to suppress hyperfine-changing (inelastic) collisions in ultracold atomic gases, as the main cause of the gas heating  \cite{zilio1996,suominen1996a}. OS was probed a powerful tool to suppress the rate of inelastic processes occurring at short distances \cite{weiner1999}. The principle of OS is displayed in Fig. \ref{fig:OSscheme}. Exposing the atoms to a continuous wave (cw) laser field detuned to the blue of a suitable atomic transition, one can drive the colliding atom pair from the initial ground state to a well-chosen electronically-excited state with a long-range repulsive potential energy curve (PEC), thus preventing the atoms to come close to each other. This ``blue-shielding'' technique is described in detail in Refs. \cite{suominen1995,napolitano1997}. 

OS between \textit{identical} atoms has been studied both experimentally and theoretically by several groups in the past decades. Walhout \emph{et al.} \cite{walhout1995} relied on the variation of the detuning of the near-resonant trapping light of a magneto-optical trap (MOT) to reduce the losses during the collision of cold metastable xenon atoms. Suominen \emph{et al.} \cite{suominen1996a} interpreted this result using several versions of semiclassical Landau-Zener (LZ) models, involving as a free parameter an average over a distribution of molecular Rabi frequencies instead of the averaged value for the molecular Rabi frequency. Further experiments have been performed with sodium atoms \cite{marcassa1995,zilio1996}, showing that OS is sensitive to the intensity and the polarization of the shielding laser. In particular, Zilio \emph{et al.}  \cite{zilio1996} demonstrated that OS is more efficient with circularly-polarized light than with linearly-polarized light. Subsequently, Napolitano \emph{et al.} \cite{napolitano1997} developed a full three-dimensional quantum scattering approach and demonstrated a qualitative agreement with the observations of Ref. \cite{zilio1996}.

The above investigations focused on OS for \textit{identical} atoms, for which the PEC for the relevant molecular excited states vary as $R^{-3}$ for large interatomic distance $R$,  due to resonant dipole-dipole interaction, some of them being repulsive and thus suitable for OS. While binary mixture of different atomic species are routinely used in ultracold physics (such as all experiments devoted to ultracold molecule formation quoted above), OS has never been explored in this context. The main difference with the case of identical atoms is that the corresponding PECs of the molecular excited states now vary as $R^{-6}$ at large distance due to van der Waals interaction, with a shorter range compared to the resonant dipole-dipole interaction.

In this work, we investigate OS between ultracold alkali-metal atoms of different species, using a Hamiltonian involving the shielding laser expressed in a basis of stationary dressed states in order to describe the coupling between electronic states induced by the laser. We numerically solve the resulting coupled-channel Schr\"{o}dinger equations to extract elastic and inelastic scattering rates. The spin-orbit coupling (SOC) in the molecular excited states is included, while the hyperfine (hf) couplings inducing inelastic collisions are simulated via an artificial loss channel. Spontaneous emission (SE), which could be a strong limiting factor for OS \cite{suominen1995}, is modeled  with an imaginary potential.

\begin{figure}[!t]\centering
\resizebox{0.45\textwidth}{!}
{\includegraphics{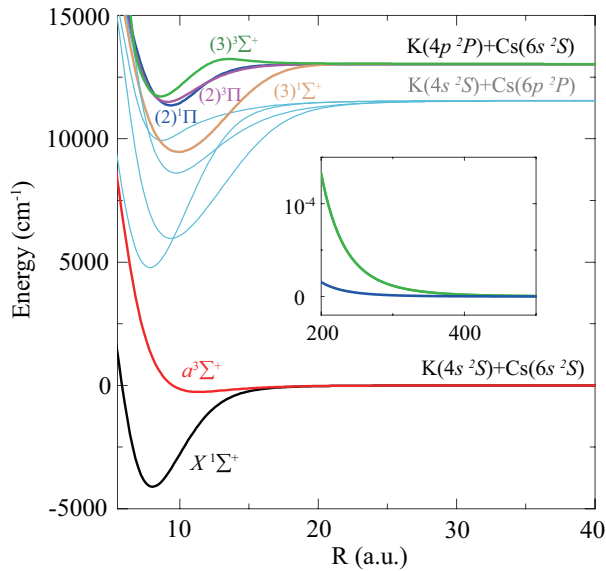}}
\caption{Hund's case {\it a} PECs of KCs relevant for the present study. The PECs of the electronic ground state $X^1\Sigma^+$ and the lowest triplet state $a^3\Sigma^+$ are taken from spectroscopic data \cite{ferber2013}. They represent the entrance channels for the atom-atom collisions. The PECs for the excited states arise from our calculations \cite{orban2019}. The inset shows that the PECs correlated to K($4p\ ^2P$)+Cs($6s\ ^2S$) limit are repulsive at large distances. Note that the (3)$^3\Sigma^+$ and (3)$^1\Sigma^+$ PECs have the same long-range coefficients, as well as the (2)$^3\Pi$ and (2)$^1\Pi$ PECs, so that only two curves (instead of four) are visible in the inset. The PECs dissociating into K($4s\ ^2S$)+Cs($6p\ ^2P$) are displayed in light blue for completeness, but are not included in the present study.}
\label{fig:PEC}
\end{figure}

Following our previous investigations \cite{orban2015,borsalino2016,orban2019}, we choose the example of the bosonic $^{39}$K-Cs mixture, which is of experimental interest \cite{groebner2017}, but still rarely explored. An overview of the PECs calculated by us, and relevant for the present study, is displayed in Fig.\ref{fig:PEC}. Their behavior at large distances is represented with a standard multipolar expansion $V_{lr}=-C_6/R^6-C_8/R^8$, taking the values of Ref.\cite{marinescu1999} for the $C_6$ and $C_8$ coefficients. As $C_6<0$ for all the states correlated to the K(4$p\ ^{2}P$)+Cs(6$s\ ^{2}S$) second excited dissociation limit, their PECs are repulsive at large distance (inset in Fig. \ref{fig:PEC}), and thus are possible candidates for OS. 

This paper is organized as follows: In Section \ref{sec:theory} the theory of OS is presented in a general atom-atom case, relying on a Hamiltonian expressed in the field-dressed-state picture. The scattering problem is formulated in this framework, paying attention to the transformation between body-fixed (BF) and space-fixed (SF) frames. We thus derive the definition for the OS efficiency, taking in account SE. In Section \ref{sec:results} we apply the model to the $^{39}$K-Cs atom pair, considering the two possible initial colliding channels $X^{1}\Sigma^{+}$ or $a^{3}\Sigma^{+}$. We discuss the shielding efficiency at ultracold temperature for linear and circular polarizations of the OS light, for a broad range of experimentally accessible shielding laser detunings $\Delta$ and Rabi frequencies $\omega$ between 0 and 200~MHz.

\section{Theory of optical shielding}
\label{sec:theory}

Our approach for cold collisions between trapped atoms assisted by light is similar to the one employed in Ref. \cite{nicholson2015} for optical Feshbach resonances. In the Born-Oppenheimer (BO) approximation, the Hamiltonian for the relative nuclear motion with coordinate $\vec{R}$ of the two alkali-metal atoms with reduced mass $\mu$ in the presence of a laser field is written as
\begin{equation}
    \begin{aligned}
    \hat{H} = -\frac{\hbar^{2}}{2\mu R^{2}}\frac{\partial}{\partial R} \left(R^2\frac{\partial}{\partial R}\right) + \frac{\hbar^{2}\hat{\ell}^{\,2}}{2\mu R^{2}}+\hat{V}(R)+\hat{V}_{\textrm{opt}}\\
    +\hat{H}_{\textrm{so}}(R)+\hat{H}_{f}+\hat{H}_{I}+\hat{V}_{\textrm{art}}.
    \label{eq:Hamiltonian}
    \end{aligned}
\end{equation}
The first term represents the relative kinetic energy operator of the two atoms about their center of mass, ${\hat{\ell}}$ is the rotational angular momentum operator of the colliding complex in the SF frame, its eigenvalue $\ell$ being associated with the partial wave. The operator $\hat{V}(R)$ represents the $R$-dependent potential energy of the system, namely the usual non-relativistic BO potential energy curves $V(R)$ of the  Hund's case $a$ electronic states $(n)^{2S+1}|\Lambda|$ of the diatomic system: $|\Lambda|$ is the projection of the total electronic orbital momentum $\vec{L}$ on the interatomic axis, $S$ is the total electronic spin, and $n$ refers to the ordering of the states with increasing energy. The optical potential $\hat{V}_{\textrm{opt}}$ is a purely imaginary operator which simulates the SE from an electronically excited BO state. The operator $\hat{V}_{\textrm{art}}$ represents an artificial channel which simulates the loss induced by the short-range dynamics in the entrance channel, related for instance to hyperfine interaction.



The spin-orbit operator $\hat{H}_{so}(R)$ is expressed in the separated-atom approximation where the total electronic orbital momentum $\vec{L}$ of the diatomic system is perfectly defined,
\begin{equation}
    \hat{H}_{\textrm{so}}(R)=A(R)\vec{L}\cdot\vec{S},
\end{equation}
where $A(R)$ specifies the spin-orbit interaction strength depending on $R$. We neglect the other similar interaction like the electronic spin-spin and spin-other-orbit interactions.

The Hamiltonian of the laser field $\hat{H}_f$ is expressed as
\begin{equation}
    \hat{H}_{f}=h\nu_L(\hat{a}_q\hat{a}^{\dag}_q-\bar{N}),
    \label{Hf}
\end{equation}
where $\nu_L$ is the frequency of the laser field with an average photon number $\bar{N}$, and $\hat{a}^{\dag}_q$ and $\hat{a}_q$ are photon creation and annihilation operators. The Hamiltonian for the one-photon interaction between the laser field and the atom pair in the electric dipole approximation is
\begin{equation}
    \hat{H}_I =-i\left(\frac{2\pi h \nu_L}{V_Q}\right)^{1/2}\vec{\mu}_M\cdot(\hat{\vec{\varepsilon}}_q a_q-\hat{\vec{\varepsilon}}_q^* a^{\dag}_q),
\end{equation}
where $V_Q$ is the quantization volume, $\vec{\mu}_M$ is the molecular electric dipole moment operator calculated in the molecular frame, and $\varepsilon_q$ denotes the components of polarization vector of the photon in the SF frame.

In order to simulate possible losses due to hyperfine interaction, we implemented an artificial channel in our coupled equations, as announced above. Experimentally the hyperfine structure (hfs) is prominent as the atoms are prepared in a given hyperfine level $F$, and even in a sublevel $M_F$ in the presence of a magnetic field. The shielding probability relies on the relative slopes of the ground and excited states. As demonstrated in \cite{orban2019} (see Fig. 4 there), in the absence of an external magnetic field, the long-range interaction between K($4P_{1/2,3/2}$) and Cs($6S_{1/2}$) including hf interactions is always repulsive regardless the populated hf level and the slopes of the potentials accounting for hfs do not change significantly with respect to the Hund's case $c$ PECs without hfs. This structure is simple for the K($4P_{1/2}$)+Cs($6S_{1/2}$) asymptote, exhibiting two isolated manifolds of non-interacting PECs very similar to those of the present model. For the K($4P_{3/2}$)+Cs($6S_{1/2}$) dissociation limit, the hfs becomes more complicated due to higher degeneracy, but remains qualitatively similar to the previous case, in particular with no significant change of the slope of the PECs at large $R$. In the presence of hfs, our model would assume that the entrance channel first interacts an excited state with the outermost repulsive PEC the hf manifold for the chosen detuning, in such a way that most of the incoming flux is sent back into the entrance channel. The part of the incoming flux which would reach shorter distances is considered as a loss, so that the next crossings with the other components of the hf manifold are not taken into account. What is mandatory here is the optimization of the shielding process with respect to the crossing between the entrance channel and the outermost component of the hfs  manifold in the excited state. A similar argument would hold in the presence of a magnetic field, namely, if the degeneracy between hyperfine levels is removed. Therefore we do not expect a significant influence of the hfs on the process.


Nicholson \emph{et al.} \cite{nicholson2015} pointed out that the \textit{uncoupled} Hund's case $e$ basis $|j,m_j\rangle|\ell,m_{\ell}\rangle$ (where $j$ and $m_j$ are the quantum numbers associated with the total electronic angular momentum of the atom pair $\vec{j}=\vec{L}+\vec{S}$, and its projection $j_z$ on the SF \emph{z}-axis) is appropriate for strong field transition problems. The $|j,m_j\rangle$ basis set is well adapted to the description of SOC in the K and Cs atoms, which is much stronger than the optical coupling. Furthermore for transition processes occurring at large distances, the quantum number $\ell$ and its projection $m_{\ell}$ on the SF \emph{z}-axis can be considered as spectator during the collision. 

The selection rules for electric dipole transitions rely on the value of the total angular momentum ${\vec J}={\vec j}+{\vec \ell}$. Thus we introduce the coupled Hund's case $e$-basis $|j,\ell,J,M_J,\epsilon\rangle$ with well-defined total parity ($\epsilon=\pm1$). According to standard notations, the parity is labeled as $e$ and $f$, corresponding to $\epsilon=(-1)^J$ and $\epsilon=(-1)^{J+1}$, respectively, for integer values of $J$, which is the case here for a two-valence-electron system \cite{julienne1984}. The average photon number $\bar{N}$ is added to the basis set, leading to the dressed coupled Hund's case $e$ basis set $|j\ell JM_J\bar{N}\epsilon \rangle$. 

When two alkali-metal atoms in their ground state $^2S$ collide (hereafter referred to as the $s+s$ case), we have $L=0$, $S=0,1$, so that $j = 0,1$. There are four relevant vectors $|j\ell JM_J\bar{N}\epsilon \rangle$ in the coupled Hund's case $e$ basis for given $J$ and $M_J$,
\begin{equation}
\begin{aligned}
    |1> &=|j=0,\ell=J,J,M_J,\bar{N},e> \\
    |2> &=|j=1,\ell=J,J,M_J,\bar{N},e> \\
    |3> &=|j=1,\ell=J-1,J,M_J,\bar{N},f> \\
    |4> &=|j=1,\ell=J+1,J,M_J,\bar{N},f>, \\
\end{aligned}
\end{equation}
where the defined parity for a given $J$ imposes the value of $\ell$. The general selection rules of electric dipole allowed (one-photon) transitions in Hund's case $e$ basis are \cite{drozdova2012}
\begin{equation}
\begin{aligned}
    &\Delta J = 0, e\leftrightarrow f(J>0) \\
    &\Delta J = \pm 1, e\leftrightarrow e,f\leftrightarrow f \\
    &\Delta M_J  = q \\
    &\Delta \bar{N} = \pm 1, \\
\end{aligned}
\end{equation}
where $q$ is the polarization of light defined in the SF frame. This restricts the number of vectors $|j'\ell' J'M'_J\bar{N'}\epsilon' \rangle$ that can be reached from the $s+s$ set above in the excited state manifold with colliding $^2S$ and $^2P$ atoms (hereafter referred to as the $s+p$ case),
\begin{equation}
\begin{aligned}
    |5\rangle &=|j',\ell',J'=J-1,M_J^{'}=M_J+q,\bar{N}-1,e\rangle \\
    |6\rangle &=|j',\ell',J'=J,M_J^{'}=M_J+q,\bar{N}-1,f\rangle \\
    |7\rangle &=|j',\ell',J'=J+1,M_J^{'}=M_J+q,\bar{N}-1,e\rangle \\
    |8\rangle &=|j',\ell',J'=J-1,M_J^{'}=M_J+q,\bar{N}-1,f\rangle \\
    |9\rangle &=|j',\ell',J'=J,M_J^{'}=M_J+q,\bar{N}-1,e\rangle \\
    |10\rangle &=|j',\ell',J'=J+1,M_J^{'}=M_J+q,\bar{N}-1,f\rangle. \\
\end{aligned}
\end{equation}
%
Here we only take into account the vectors with $\bar{N}-1$ as it will be clarified further below. The specific values of $\ell'$ in the above expressions are not given since they are related to $j'=0,1,2$ and $J$ in terms of algebra of angular momentum algebra.

We now derive the matrix elements of the Hamiltonian of Eq.\ref{eq:Hamiltonian} in the dressed Hund's case $e$ coupled basis set. The centrifugal part is diagonal,
\begin{equation}
\begin{aligned}
     \langle j\ell JM_J\bar{N}\epsilon|\frac{\hbar^{2}\hat{\ell}^{\,2}}{2\mu R^{2}}|j'\ell'J'M_J'\bar{N}'\epsilon'\rangle \\
     =\delta_{j j'}\delta_{\ell \ell'} \delta_{J J'} \delta_{M_J M'_J} \delta_{\bar{N} \bar{N}'} \delta_{\epsilon  \epsilon'} \frac{{\hbar}^2\ell(\ell+1)}{2\mu R^2}.
\end{aligned}
\end{equation}
%
The  matrix elements of $\hat{V}(R)$ and $\hat{H}_{so}(R)$ are first calculated in the dressed Hund's case $a$ basis $|^{2S+1}\Lambda\Sigma|\Omega|JM_J\bar{N}\epsilon\rangle$ combining the BO electronic states $|a_{\textrm{BO}}\rangle \equiv |^{2S+1}\Lambda\Sigma|\Omega|\rangle$ to rotational functions to form total wave function with a good total parity $\epsilon$ \cite{singer1983},
\begin{equation}
\begin{aligned}
   & |^{2S+1}\Lambda\Sigma|\Omega|JM_J\bar{N}\epsilon\rangle \\ 
   &= \sqrt{\frac{1}{2(1+\delta_{\Lambda,0}\delta_{\Sigma,0})}}\frac{\sqrt{2J+1}}{4\pi}\left[D_{M_J\Omega}^{J*}(\alpha,\beta)|\Lambda S\Sigma\rangle \right.\\ &\left. +\epsilon(-1)^{J-S+\sigma}D_{M_J-\Omega}^{J*}(\alpha,\beta)|-\Lambda S-\Sigma\rangle \right]|\bar{N}\rangle,
 \label{eq:hunda}
\end{aligned}
\end{equation}
where $\Omega$ = $\Lambda$+$\Sigma$. We recall that in this basis vector the quantum number $\Sigma$ is different from the character `$\Sigma$' labeling Hund's case $a$ electronic states with $\Lambda=0$. The Wigner functions $D_{M_J\Omega}^{J*}$ and $D_{M_J-\Omega}^{J*}$ are the rotation matrix elements. The Euler angles ($\alpha,\beta,\gamma$) characterize the rotation from the SF to the BF frames. Finally,  $\sigma=0$ for all Hund's case $a$ molecular symmetries but the $\Sigma^{-}$ symmetry (not involved here) with $\sigma=1$. 

By using the orthogonal frame transformation $U_{e\rightarrow a}$ from Hund's cases $e$ to $a$ defined in Ref.\cite{singer1983,dubs1991}, we obtain the matrix elements of $\hat{V}+\hat{V}_{\textrm{opt}}$ in the dressed Hund's case $e$ coupled basis,
\begin{equation}
\begin{aligned}
    &\langle j\ell JM_J\bar{N}\epsilon|\hat{V}+\hat{V}_{\textrm{opt}}|j'\ell'J'M_J'\bar{N}'\epsilon'\rangle \\
    &=U_{e\rightarrow a}\langle a_{\textrm{BO}}| \hat{V}+\hat{V}_{\textrm{opt}}| a'_{\textrm{BO}}\rangle U_{e\rightarrow a}^T\\
    &=U_{e\rightarrow a}\left\{\left[
    V(R)-i\hbar\Gamma/2\right]\delta_{a_{\textrm{BO}} a'_{\textrm{BO}}}\right\}U_{e\rightarrow a}^T,
    \label{eq:v_ele}
\end{aligned}
\end{equation}
We have introduced the purely imaginary optical potential $V_{\textrm{opt}}= -i\hbar\Gamma/2$ to simulate the SE process, with the natural line width $\Gamma$ of the relevant radiative atomic transition. It is set to 0 when $R > 500$~a.u. to emphasize on SE within the interaction region. The matrix elements of $U_{e\rightarrow a}$ are expressed as \cite{singer1983,bergeman2002}
\begin{equation}
\begin{aligned}
    &\langle j\ell JM_J\bar{N}\epsilon|^{2S+1}\Lambda\Sigma|\Omega|JM_J\bar{N}\epsilon\rangle \\
    &=\frac{1+\epsilon(-1)^{\ell+L_A+L_B}}{\sqrt{2(1+\delta(\Sigma,0)\delta(\Lambda,0))}}(-1)^{\ell+\Omega-J}\\
    &\times\sqrt{2S+1}\sqrt{2j_A+1}\sqrt{2j_B+1}\langle j,J,-\Omega,\Omega|\ell,0\rangle\\
    &\times\sum_{L\Lambda_A\Lambda_B}\sqrt{2L+1}\langle L_A,L_B,\Lambda_A,\Lambda_B|L,\Lambda\rangle\\
    &\times\langle L,S,\Lambda,\Sigma|j,\Omega\rangle\left\{ \begin{array}{ccc}
                                                            L_A & s_A & j_A \\
                                                            L_B & s_B & j_B \\
                                                            L   &  S  & j
                                                            \end{array}\right\}
    \langle\Lambda_A,\Lambda_B|c,\Lambda\rangle,
\end{aligned}
\end{equation}
where $L_{A(B)}$ and $s_{A(B)}$ specify the quantum numbers of electronic orbital and spin angular momenta of atom $A$($B$), respectively. $\Lambda_{A(B)}$ is the projection of $\vec{L}_{A(B)}$ on the interatomic axis and $\vec{j}_{A(B)}$ = $\vec{L}_{A(B)}$ + $\vec{s}_{A(B)}$. The expressions, such as $\langle j,J,-\Omega,\Omega|\ell,0\rangle$, denote the Clebsch-Gordan coefficients and $\{\cdots\}$ is 9$j$ coefficient.  The last term in the symbol $\langle$ $|$ $\rangle$ denotes the coupling at $R\rightarrow +\infty$ in the $L_AL_B\Lambda$ threshold, c labels different states converging towards the same dissociation limit $L_A + L_B$, with the same $\Lambda$ value but different couples ($\Lambda_A , \Lambda_B)$. For $p$+$s$ limit $L_B$ = 0 and $\Lambda_B$ = 0 hence $\Lambda_A$ = $\Lambda$ and $\langle\Lambda_A,\Lambda_B|c,\Lambda\rangle$ = 1 \cite{singer1983,bergeman2002}. 

Similarly, the $\hat{H}_{\textrm{so}}(R)$ matrix elements are
\begin{equation}
\begin{aligned}
     &\langle j\ell JM_J\bar{N}\epsilon|\hat{H}_{\textrm{so}}(R)|j'\ell'J'M_J'\bar{N}\epsilon'\rangle \\
     &=U_{e\rightarrow a}\langle a_{\textrm{BO}}| A(R)\vec{L}\cdot\vec{S} |a'_{\textrm{BO}}\rangle U_{e\rightarrow a}^T.
\end{aligned}
\end{equation}
The matrix elements of $\vec{L}\cdot\vec{S}$ relevant to $s+p$ dissociation limit in Hund's case $a$ basis can be found in Ref.\cite{orban2015}. 

The laser field Hamiltonian $\hat{H}_f$ is diagonal in dressed Hund's case $e$ basis,
\begin{equation}
\begin{aligned}
    &\langle j\ell JM_J\bar{N}\epsilon|\hat{H}_f|j'\ell'J'M_J'\bar{N}\epsilon'\rangle \\
    & = \delta_{j j'}\delta_{\ell \ell'} \delta_{J J'} \delta_{M_J M'_J} \delta_{\bar{N} \bar{N}'} \delta_{\epsilon  \epsilon'} h\nu_L(\bar{N}+1/2).
\end{aligned}
\end{equation}
The $\hat{H}_I$ matrix elements were displayed in Refs.  \cite{napolitano1997,napolitano1998} in the dressed Hund's case $e$ uncoupled basis,
\begin{equation}
\begin{aligned}
   &\langle jm_j\ell m_\ell \bar{N}\epsilon|\hat{H}_I|j'm_j'\ell'm_\ell'\bar{N}'\epsilon'\rangle \\
   &=\hbar\omega_{j\rightarrow j'}\delta_{m_j,m_j'+q}\,\delta_{\ell,\ell'}\, \delta_{m_\ell,m_\ell'}\, \delta_{\bar{N},\bar{N}'\pm1}\, \delta_{\epsilon,-\epsilon'},
\end{aligned}
\end{equation}
where $\omega_{j\rightarrow j'}$ denotes the molecular Rabi frequency which in principle can be extracted from the measured oscillator strength of the relevant transition, which includes dependency on $j$ and $j'$ through H\"onl-London factors \cite{julienne1984}. For simplicity sake, we ignore these factors so that $\omega_{j\rightarrow j'} \equiv \omega$, $\omega$ being the generic Rabi frequency used afterwards. The matrix elements in the dressed Hund's case $e$ coupled basis are obtained through summations over the projections  $m_j$ and $m_\ell$ on the space fixed $z$-axis,
\begin{equation}
\begin{aligned}
   & \langle j\ell JM_J\bar{N}\epsilon|\hat{H}_I|j'\ell' J'M_J'\bar{N}-1\epsilon'\rangle \\
   & =\sum_{m_j,m_j'}\sum_{m_\ell,m_\ell'}\langle jm_j\ell m_\ell|JM_J\rangle\langle j'm_j'\ell'm_\ell'|J'M_J'\rangle \\
   & \times\langle jm_j\ell m_\ell \bar{N}\epsilon|\hat{H}_I|j'm_j'\ell'm_\ell'\bar{N}-1\epsilon'\rangle.\\
\end{aligned}
\label{eq:Hi_transf}
\end{equation}
From Eq. \ref{eq:Hi_transf} it is useful to recast the selection rules $\Delta J=0,\pm1$ for a dipole-allowed transition, correlated to the change of total parity $\epsilon'=-\epsilon$. For $e$-levels (resp. $f$-levels) we have  $\epsilon (-1)^J= +1$ (resp.  $\epsilon (-1)^J= -1$). Then,
\begin{equation}
\begin{aligned}
&\epsilon \epsilon' (-1)^{J+J'}=(-1)^{J+J'+1}&\\
=&\begin{cases}
+1,\, {\rm for\,\,} e\rightarrow e,\,\,{\rm and}\,\, f\rightarrow f & {\rm leading\, to\, \Delta J=\pm 1,}\\
-1,\, {\rm for}\,\, e\rightarrow f & {\rm leading\, to\, \Delta J=0.}
\end{cases}
\end{aligned}
\label{eq:selrules}
\end{equation}
The incoming flux loss induced by the short-range dynamics is taken into account through an artificial potential operator $\hat{V}_{\textrm{art}}$, following \cite{napolitano1997,julienne1994}. It is diagonal in the dressed Hund's case $a$ basis, and the corresponding PEC is chosen to cross the $X^1\Sigma^+$ or $a^3\Sigma^+$ PECs of the entrance channels at their inner wall, with an arbitrary shape. The corresponding coupling must be tiny in order to minimize the influence of the artificial channel on the final results. The absolute flux loss probed by the artificial channel is meaningless and we are only interested in its variation with the Rabi frequency and the detuning of the shielding laser. We choose the expression
\begin{equation}
    V_{\textrm{art}}(R) = C_{\textrm{art}}/R^3-\hbar\Delta_{\textrm{art}},
\label{eq:artificial}
\end{equation}
depending on two parameters $C_{\textrm{art}}$ and $\Delta_{\textrm{art}}$ specified further down.

We expand the total scattering wave function $|\Psi(R)\rangle$ of the diatomic system on the dressed Hund's case $e$ coupled basis in the SF frame,
\begin{equation}
|\Psi(R)\rangle=\sum_{\alpha}\frac{F_{\alpha}(R)}{R}|\alpha\rangle,
\label{eq:wvf}
\end{equation}
where $F_{\alpha}(R)$ is the radial wavefunction associated to the basis vector $|\alpha\rangle$ = $|j\ell JM_J{\bar N}\epsilon\rangle$. Substituting Eq. (\ref{eq:Hamiltonian}) and Eq. (\ref{eq:wvf}) in the time-independent Schr\"{o}dinger equation $\hat{H} \Psi = E_{col} \Psi$ results in a set of coupled-channel (CC) equations for the collision energy $E_{col}$ with the general form
\begin{equation}
\begin{aligned}
\left[-\frac{\hbar^2}{2\mu}\frac{d^2}{dR^2}+\frac{{\hbar}^2\ell(\ell+1)}{2\mu R^2}-E_{col}\right]F_{\alpha}(R) \\
 =  -\sum_{\alpha'}\langle\alpha|\hat{V}(R)+\hat{V}_{\textrm{opt}}+ \hat{H}_{\textrm{so}}(R)+\hat{H}_f+\hat{H}_I(R)\\
+\hat{V}_{\textrm{art}}|\alpha'\rangle F_{\alpha'}(R).
\end{aligned}
\end{equation}
The CC equations are numerically solved with the log-derivative method \cite{johnson1973} involving a complex optical potential \cite{tuvi1993}. The $S$-matrix is extracted at large distances after matching the general boundary conditions to the log-derivative matrix. The scattering boundary conditions are written in a basis resulting from the diagonalization of ${\hat H}$ at infinity \cite{napolitano1997}.

\begin{figure}[!t]\centering
\resizebox{0.45\textwidth}{!}
{\includegraphics{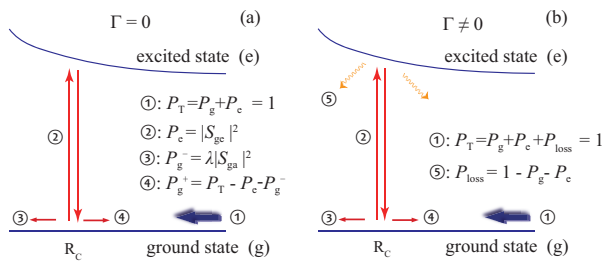}} 
\caption{Illustration with a two-channel case (in an undressed picture, for clarity) of the various probabilities relevant for the definition of the OS efficiency, related to various elements $S_{xy}$ of the scattering matrix $S$ for a collision energy $E_{col}$ and a Rabi frequency $\omega$. (a) neglecting spontaneous emission ($\Gamma=0$); (b)including spontaneous emission ($\Gamma \neq 0$). In (a) the total probability $P_T=1$ entering the incoming channel $g$ splits in the probabilities $P_e=|S_{ge}|^2$ to reach the excited channel $e$ due to the OS laser, $P_g^-=\lambda |S_{ga}|^2$ to react in the inner zone from the $g$ channel and ''detected'' by the artificial channel $a$ with a scaling factor $\lambda$, and $P_g^+=P_T-P_e-P_g^- \equiv |S_{gg}|^2$ to be reflected back into the entrance channel. When including the spontaneous emission in (b), an additional loss probability $P_{loss}$ must be considered, so that $P_T=P_g^++P_g^-+P_e+P_{loss} \equiv 1$.
}
\label{fig:proba}
\end{figure}

Figure \ref{fig:proba}(a) illustrates the way to derive the shielding efficiency from the computed $S$-matrix. Assuming a two-channel case, and in the absence of SE, Napolitano \cite{napolitano1997} introduced the shielding parameter $P_S^{no\,SE}$: it expresses the modification of the flux reacting at short distances and absorbed by the artificial channel due to the OS laser, as the ratio between the probability to react in the inner zone $P_g^-$ in the presence of the OS laser and without it, 
\begin{equation}
    \begin{aligned}
    P_S^{no\,SE}=\frac{P_g^-(E_{col},\omega)}{P_g^-(E_{col},0)} \equiv \frac{|S_{ga}(E_{col},\omega)|^2}{|S_{ga}(E_{col},\omega=0)|^2}.
    \end{aligned}
\end{equation}
It is independent from the scaling parameter $\lambda$, assuming that it does not vary with the Rabi frequency $\omega$. Thus the shielding probability is $1-P_S^{no\,SE}$, \textit{i.e.} $P_S^{no\,SE}\ll$ 1 expresses efficient OS. Without loss of generality, we take into account the contributions from high partial waves, and the shielding parameter can be expressed as
\begin{equation}
    \begin{aligned}
    P_S^{no\,SE}=\frac{\sum_{\ell\ell'}|S_{ga}^{\ell\ell'}(E_{col},\omega)|^2}{\sum_{\ell\ell'}|S_{ga}^{\ell\ell'}(E_{col},\omega=0)|^2},
    \label{eq:ps}
    \end{aligned}
\end{equation}
where $S_{ga}^{\ell\ell'}$ characterizes the $S$ matrix element between $g$ with partial wave $\ell$ and $a$ with $\ell'$.

Spontaneous emission (Figure \ref{fig:proba}(b)) is expected to induce an atom loss from the trap, as the emitted photon may have an energy significantly different from the absorbed one, yielding additional kinetic energy to the atoms. In the formalism, the $S$ matrix is not unitary anymore, and SE generates a probability for loss of flux in the entrance channel $P_{loss}=1-P_g^+-P_g^--P_e$, which is thus related to the imaginary part of the $S$-matrix: $P_{loss}=1-|S_{gg}|^2-|S_{ge}|^2$. The expression of the shielding parameter $P_S$ remains identical to Eq. \ref{eq:ps}, except that the $S$ matrix has now changed.

In the next section we explore the variation of $P_S$ and $P_{loss}$, as functions of the Rabi frequency $\omega$ and the detuning $\Delta$. We also define the elastic $k_{el}=\sigma_{el}(2E_{col}/\mu)^{1/2}$ and the inelastic  $k_{in}=\sigma_{in}(2E_{col}/\mu)^{1/2}$ collision rate coefficients, where the cross sections $\sigma_{el}$ and $\sigma_{in}$ are determined, in the context of the two-channel picture of Figure \ref{fig:proba}, by the $S$-matrix elements $S_{gg}$ and $S_{ge}$, respectively.

\section{Results and Discussion}
\label{sec:results}

We consider the ultracold collision of $^{39}$K-Cs in their ground states. The Hund's case $a$ PECs for the $X^1\Sigma^+$ and $a^3\Sigma^+$ states correlated to $^{39}$K(4s)+Cs(6s) are taken from the spectroscopic investigation of Ref.\cite{ferber2013}. The PECs for the excited states dissociating to K($4p$)+Cs($6s$) are calculated by us up to $R = 40 $~a.u. following the methodology described in Ref.\cite{aymar2005,aymar2006,orban2015,borsalino2016,orban2019}. They are extrapolated with the multipolar expansion $V_{lr}=-C_6/R^6-C_8/R^8$ using the coefficients of Ref. \cite{marinescu1999}: $C_6(\Sigma) = -3.8865 \times 10^4$~a.u., $C_8(\Sigma) = 8.6801 \times 10^6$~a.u., $C_6(\Pi) = -4.5843 \times 10^3$~a.u., and $C_8(\Pi) = 1.7915 \times 10^6$~a.u.. 

There are no published molecular $R$-dependent SOC strengths $A(R)$ for the K($4p$)+Cs($6s$) dissociation limit. As we focus on interactions at large distance, we can safely adopt a constant SOC based on the atomic spin-orbit splitting $\Delta E^{\textrm{K}}_{so}=57.71$~cm$^{-1}$ of K($4p)$ \cite{NIST_ASD}: $A(R)=\Delta E^{\textrm{K}}_{so}/3$. The radiative decay half-width for K($4P_{3/2}$) and K($4P_{1/2}$) are  $\Gamma/2\pi=6.035$~MHz and 5.956~MHz \cite{wang1996}.

\begin{figure}[!t]\centering
\resizebox{0.45\textwidth}{!}
{\includegraphics{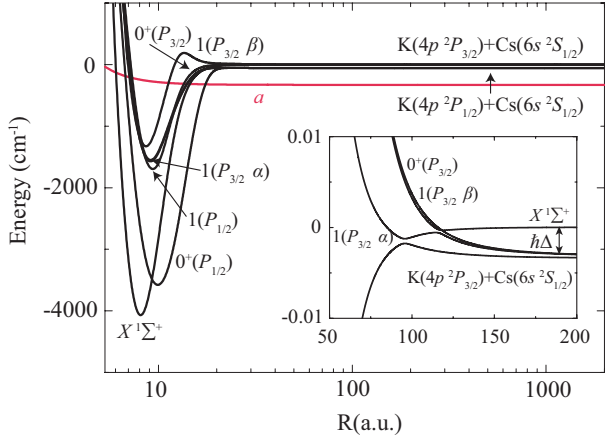}} 
\caption{The dressed adiabatic Hund's case $e$ PECs in the SF frame, setting $J = 0$ in the $X^1\Sigma^+$ entrance channel with $\bar{N}$ photons, coupled with a linearly-polarized shielding laser to $J' = 1$ levels in the electronically-excited states correlated to  $^{39}$K$(4\,^2P_{3/2}$+Cs$(6\,^2S_{1/2}$ with $\bar{N}-1$ photons. The values of the Rabi frequency $\omega = 10$~MHz,  detuning $\Delta = 100$~MHz are chosen to clearly exhibit the avoided crossings induced by the OS laser (inset). At infinity, the PECs for the incoming channel and the excited channels are separated by the energy $\hbar \Delta$. Only $e$-parity states are displayed for clarity, consistently with the selection rule for a $J\rightarrow J+1$ transition (Eq. \ref{eq:selrules}). The excited PECs are labeled with their dominant Hund's case $c$ character, with the indexes $\alpha$ and $\beta$ distinguishing the two 1($P_{3/2}$) states. The PEC of the artificial channel (red curve) is labeled with \emph{a}. A figure for the $a^3\Sigma^+$ entrance channel coupled to $f$-parity states is provided in the Appendix.}
\label{fig:dressed}
\end{figure}

The parameters of the artificial potential $V_{\textrm{art}}(R)$ of Eq.\ref{eq:artificial} are chosen as follows: $C_{\textrm{art}} = 0.2$~a.u., and $\Delta_{\textrm{art}} = 0.0015$~a.u. (resp. $\Delta_{\textrm{art}} = 0.0005$~a.u.) for the $X^1\Sigma^+$ (resp. $a^3\Sigma^+$) entrance channel. The latter defines its relative energy position with respect to the one of the entrance channel. It crosses the entrance channels for $R <$ 10 a.u. (Fig. \ref{fig:dressed}), and we fix the coupling between $g$ and $a$ channels to $\hbar \times $10~MHz, \textit{i. e.} smaller than the typical range for the Rabi frequency and detuning discussed below. 

In the following, only ground and excited channels coupled by only one photon must be considered. Indeed, in heteronuclear pairs, the energy of the entrance channel with two atoms in their ground state dressed by two resonant photons will be far from any other excited channel. This is in contrast with homonuclear atomic pairs, where the energy of a pair of excited atoms will be close to the one of two-photon-dressed entrance channel. For the OS between Na atoms, this circumstance has been actually used for the detection of OS \cite{zilio1996}. Entrance channels for collision between ultracold polar molecules dressed with several microwave photons have also to be considered in mw-shielding \cite{karman2018,lassabliere2018}. In this respect the present proposal is simpler than the other proposals.

An example of the one-photon-dressed adiabatic PECs in the SF frame is displayed in Figure \ref{fig:dressed}, for $J=0$, and $J'=1$ in the excited channels correlated to $^{39}$K$(4\,^2P_{3/2})$+Cs$(6\,^2S_{1/2})$, with a linearly-polarized OS laser with $\omega=10$ MHz and $\Delta=200$ MHz. Under such conditions, the avoided crossing between dressed channels are clearly visible. Thus only $e$-parity states are relevant, according to Eq.\ref{eq:selrules}, and the entrance channel is $X^1\Sigma^+$ state as only $s$-wave is allowed here. The OS laser frequency is chosen close to the one of the $^{39}$K D2 line (see the inset). Due to the repulsive long-range behavior of the Hund's case $a$ excited PECs quoted above \cite{marinescu1999}, the Hund's case $c$ excited PECs are all repulsive as well.



The OS efficiency results from the competition of several factors, as exemplified by previous studies in the homonuclear case: (i) the steepness of the excited channel PEC at the crossing point with the PEC of the dressed entrance channel which tends to minimize SE if this steepness is large enough \cite{suominen1995}; (ii) a significant light-induced coupling, or Rabi frequency $\omega$, as demonstrated by the Landau-Zener and wave packet dynamics approaches \cite{suominen1995,suominen1996a}; (iii) the laser detuning $\Delta$ of the OS laser frequency with respect to the chosen resonant atomic transition, large enough to minimize SE, while keeping a noticeable oscillator strength to limit the required OS laser intensity.


\begin{figure}[!t]\centering
\resizebox{0.5\textwidth}{!}
{\includegraphics{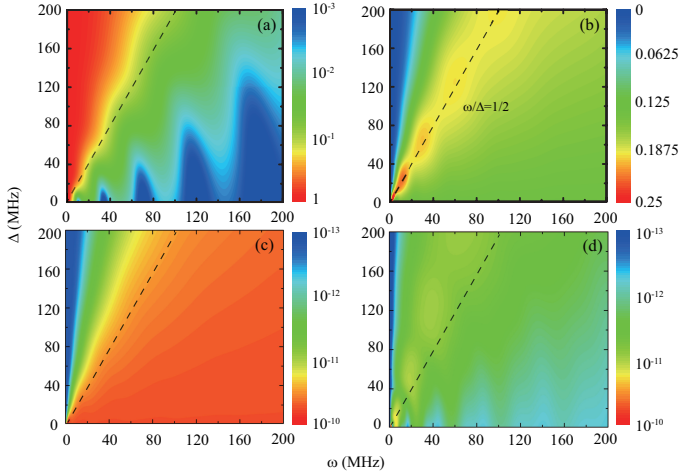}} 
\caption{Colored-contour plot of (a) $P_S$, (b) SE loss probability $P_{loss}$, (c) elastic and (d) inelastic collision rate coefficient (in cm$^3/$s), with the Rabi frequency $\omega$ and the detuning $\Delta$, of a linearly-polarized OS laser, and a collision energy $E_{col}=k_B \times 240$~nK in the entrance channel $X^1\Sigma^+$ of $e$ parity coupled to the states correlated to K$(4p\ ^2P_{3/2})$ + Cs$(6s\ ^2S_{1/2})$.} 
\label{fig:rates}
\end{figure}

Figure \ref{fig:rates} displays the colored-contour plots for $P_S$, $P_{loss}$, $k_{el}$ and $k_{in}$ in the $(\omega,\Delta)$ plane, for $E_{col}=k_B \times 240$~nK, for a linearly-polarized OS laser. Values for the total angular momentum quantum number $J = 0,1,2$ (in the entrance channel) and $J' = 0,1,2,3$ (in the excited channel) are introduced in calculations to ensure the convergence of the results. The short-range dynamics is strongly suppressed for $\omega/\Delta>1/2$ as the shielding parameter $P_S$ is well below $10^{-1}$ (panel a). In the same region, the inelastic rate $k_{in}$ is always dominated by $k_{el}$, so that the overall shielding efficiency is satisfactory. We note that the rate coefficient $k_{in}$ exhibits marked oscillations in the region $\omega/\Delta>1$, \textit{i.e.} when $k_{in} \ll k_{el}$ (Fig. \ref{fig:rates}d). These oscillations result from the interference between incoming and outgoing fluxes travelling along the two possible paths generated by the light-induced avoided crossings in Fig. \ref{fig:dressed} (also known as St\"uckelberg oscillations in the context of the semi-classical model for an avoided crossing). For the same reason these oscillations are present with the same amplitude in $k_{el}$ (Fig. \ref{fig:rates}c) but are barely visible as they lie upon a quantity with a much larger magnitude than $k_{in}$. Oscillations also occur in $P_S$ (Fig. \ref{fig:rates}a) as the numerical manifestation of the crossing of the PECs of the entrance channel and the artificial channel at short distances, but then they are physically irrelevant, as the artificial channel in itself.

The SE-induced loss probability $P_{loss}$ is minimized when $\omega/\Delta$ is far from 1/2 (see the dashed line in panel b). It reaches an almost constant value of 0.125 when $\omega/\Delta>1$, namely, in the region where OS is efficient. However this probability corresponds to a rate coefficient $k_{loss}=P_{loss}\pi\hbar/(\mu k_B)= 7.66 \times 10^{-10}$~cm$^3/s$, thus stating that SE remains an important limiting process for OS in this range, according to this stationary model based on an imaginary optical potential. One possibility to overcome this limitation is to further increase the detuning of the OS laser frequency further into the blue, while keeping $\omega/\Delta>1$, which has been demonstrated experimentally and supported by quantum Monte-Carlo simulations \cite{suominen1995,suominen1996b}. Following Ref.\cite{nicholson2015}, it is likely that for such an ultracold collision, only a fraction of the excited atoms indeed leaves the trap, thus reducing the impact of SE. Such atoms could then be implied in an atomic pair reabsorbing a photon, but this cannot be taken into account in the present time-independent approach.




While the discussion and figures above only involve the $X^1\Sigma^+$ entrance channel, we have also included the $a^3\Sigma^+$ entrance channel, which leads to both $e$ and $f$ parities in the SF frame. In the excited manifold, the $0^-(P_{3/2})$ state, resulting from the SOC between the $(2)^3\Sigma^+$ and $(1)^3\Pi$ Hund's case $a$ states, must be included too in the calculations. We display in the Appendix a figure similar to Figure \ref{fig:dressed} for the triplet entrance channel.
In Table \ref{tab:rates} we report the shielding parameter $P_S$, the SE-induced loss probability $P_{loss}$, and the elastic and inelastic rate coefficients $k_{el}$ and $k_{in}$ for both entrance channels $X^1\Sigma^+$ and $a^3\Sigma^+$, both parities $e$ and $f$, and for an OS laser with $\omega=200$~MHz and a frequency blue-detuned from the D2 line of $^{39}$K by $\Delta=200$~MHz. This corresponds to the upper right corner of the four panels in Fig.\ref{fig:rates}, \textit{i.e.} an acceptable compromise for OS efficiency, as previously argued. The Table only implies $M_J = 0$ as $s$-wave collisions are dominant at the chosen collision energy. Non-zero $M_J$ only appears for $\ell>0$ partial-waves, which have little influence on the final results. This is immediately visible in Table \ref{tab:rates} when the lowest partial wave $\ell_{\textrm{min}}$ involved in the collision is larger than 0, yielding very small rate coefficients. The overall data of the Table confirm the trends visible in Fig. \ref{fig:rates}: the OS is quite efficient ($P_S<3$\%), with an elastic rate coefficient dominating the inelastic one for both polarizations of the OS laser. The efficiency is similar for both singlet and triplet entrance channels as the corresponding electronic transition dipole moments have magnitudes comparable to each other \cite{orban2019}. This is in striking contrast with OS for identical atoms \cite{zilio1996}, or microwave shielding with identical polar molecules \cite{karman2018,lassabliere2018}, as in both cases the particle permutation operation imposes selection rules for state coupling. This condition is relaxed for OS with different atoms, or OS with identical polar molecules as the later involve two different molecular electronic states \cite{xie2020}. We note however a few exceptions in Table \ref{tab:rates} (lines 3, 11, 12) for which OS is not efficient ($P_S \approx 1$), due to the absence of coupling of the entrance channel with the excited ones, as commented further below. They all concern OS with respect to the D1 line, so that the preference should go to the D2 line for an efficient OS implementation.

\begin{table}[!t]
    \centering
    \begin{tabular}{@{}c|c|c|c|c|c|c|c|c@{}}
 \hline
              &$\epsilon$&$|q|$& $g$ & $\ell_{\textrm{min}}$ & $P_{S}$ & $P_{loss}$ & $k_{el}$ & $k_{in}$   \\ 
              &          &     &    &                       &          &            &(cm$^3/$s)&(cm$^3/$s)\\ \hline
    $P_{3/2}$ & \emph{f} & 1 & T & 0 & 0.0276 & 0.1725 & 4.70$\times10^{-11}$ & 1.69$\times10^{-12}$\\
    $P_{3/2}$ & \emph{f} & 0 & T & 0 & 0.0294 & 0.1727 & 4.71$\times10^{-11}$ & 1.70$\times10^{-12}$\\
    $P_{1/2}$ & \emph{f} & 1 & T & 0 & 0.9958 & 0.0008 & 2.74$\times10^{-12}$ & 2.02$\times10^{-15}$\\
    $P_{1/2}$ & \emph{f} & 0 & T & 0 & 0.0188 & 0.1753 & 4.70$\times10^{-11}$ & 1.01$\times10^{-13}$\\    \hline
    $P_{3/2}$ & \emph{e} & 1 & S & 0 & 0.0096 & 0.1709 & 4.73$\times10^{-11}$ & 2.15$\times10^{-12}$\\
    $P_{3/2}$ & \emph{e} & 0 & S & 0 & 0.0096 & 0.1709 & 4.73$\times10^{-11}$ & 2.15$\times10^{-12}$\\
    $P_{3/2}$ & \emph{e} & 1 & T & 1 & 0.0062 & 0.0015 & 4.61$\times10^{-16}$ & 5.60$\times10^{-15}$\\
    $P_{3/2}$ & \emph{e} & 0 & T & 1 & 0.0051 & 0.0011 & 4.82$\times10^{-16}$ & 5.38$\times10^{-15}$\\
    $P_{1/2}$ & \emph{e} & 1 & S & 0 & 0.0069 & 0.1740 & 4.13$\times10^{-11}$ & 5.88$\times10^{-13}$\\
    $P_{1/2}$ & \emph{e} & 0 & S & 0 & 0.0069 & 0.1740 & 4.13$\times10^{-11}$ & 5.88$\times10^{-13}$\\
    $P_{1/2}$ & \emph{e} & 1 & T & 1 & 0.0049 & 0.0009 & 3.02$\times10^{-16}$ & 7.85$\times10^{-16}$\\
    $P_{1/2}$ & \emph{e} & 0 & T & 1 & 1.0000 & 0      & 5.40$\times10^{-16}$ & 0\\
   \hline
    \end{tabular}
    \caption{Values of $P_{S}$, $P_{loss}$, $k_{el}$, and $k_{in}$, for the entrance channel $g$ set to $X^1\Sigma^+$ (S) and $a^3\Sigma^+$ (T) with parity $\epsilon = e,f$, $M_J = 0$, involving a lowest partial wave $\ell_{\textrm{min}}$ at the collision energy $E_{col}=k_B \times 240$~nK. The OS laser polarization is either linear ($|q|=0$) or circular ($|q|=1$), with a Rabi frequency $\omega = 200$~MHz and a detuning $\Delta = 200$~MHz from the D2 (labeled $P_{3/2}$) or the D1 (labeled $P_{1/2}$) $^{39}$K line.}
    \label{tab:rates}
\end{table}

\begin{figure}[!t]\centering
\resizebox{0.48\textwidth}{!}
{
  \includegraphics{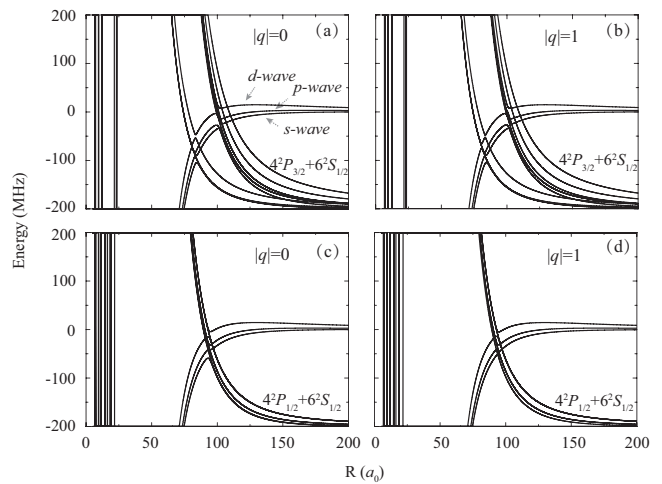}
} 
\caption{The dressed adiabatic Hund's case $e$ PECs in the SF frame, considering the $a^3\Sigma^+$ state with $f$ parity as the entrance channel. Values of $J$ up to 2 are considered so $s$, $p$ and $d$-waves are present in the calculations. The values of the Rabi frequency $\omega = 10$~MHz, and detuning $\Delta = 200$~MHz are chosen to clearly exhibit the avoided crossings induced by the OS laser. (a) Linearly-polarized ($|q|=0$) OS laser blue-detuned from the D2 transition in $^{39}$K; (b) same with $|q|=1$; (c) Linearly-polarized ($|q|=0$) OS laser blue-detuned from the D1 transition in $^{39}$K; (d) same with $|q|=1$.
} 
\label{fig:DPECS10}
\end{figure}

Another reason for the weak polarization dependence of the OS with different alkali-metal atoms can be found in Fig.\ref{fig:DPECS10}, displaying the dressed adiabatic Hund's case $e$ PECs with $f$ parity corresponding to the first four lines of Table~\ref{tab:rates}. As all excited molecular PECs are repulsive at large distance (both $\Sigma$ and $\Pi$ Hund's case $a$ PECs are repulsive too, see Fig. \ref{fig:PEC}), the PECs for the entrance channel crosses all possible excited PECs. In contrast, for the lowest $s+p$ excited PEC manifold of identical alkali-metal atoms, the Hund's case $a$ $\Sigma$ PECs are attractive while the $\Pi$ PECs are repulsive. The former cannot be crossed by the entrance channel PEC in the OS configuration, thus inducing an additional selection rule for OS.

\begin{figure}[!t]\centering
\resizebox{0.48\textwidth}{!}
{
  \includegraphics{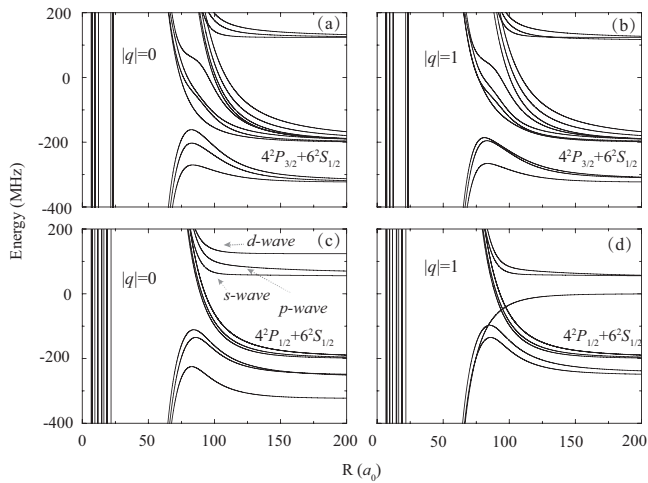}
} 
\caption{Same as Fig. \ref{fig:DPECS10} for $\omega = 200$~MHz and $\Delta = 200$~MHz.} 
\label{fig:DPECS200}
\end{figure}

Figure \ref{fig:DPECS10} is helpful to understand the few exceptions noticed in Table \ref{tab:rates}. The one at line 3 is related to panel (d) with circularly-polarized light: we see that the entrance PEC in the $s$-wave has no avoided crossing with the excited PECs, and thus OS is almost inefficient. The next partial waves are coupled but their contribution is minor, as already quoted. In contrast, panel (c) with linear polarization shows real avoided crossings, thus suitable for OS. This is even clearer in Fig. \ref{fig:DPECS200}, showing the same PECs than those of Fig. \ref{fig:DPECS10} , but with a large Rabi frequency $\omega = 200$~MHz. The light-induced avoided crossings are now hardly visible as $\omega$ has the same magnitude than $\Delta$. The same explanation still holds in the $e$ parity cases of lines 11 and 12 of Table~\ref{tab:rates}.

Finally, the issue of the detection of OS must be addressed. In the experiments dealing with ultracold polar molecules, in which an anomalously short and yet unexplained lifetime of the molecules in the trap is recorded \cite{ospelkaus2010a,ni2010,takekoshi2014,molony2014,guo2016,gregory2019}, OS would be obviously detected through an increase of this lifetime. In the original experiment of Ref.\cite{zilio1996} with Na atom pairs, a unique circumstance was used: the two-photon excited channel Na($3p\,^2P$)+Na($3p\,^2P$) has a larger energy than the one of the lowest bound level of Na$^+_2$, so that this channel undergoes autoionization. Thus OS was detected through a reduction of the autoionization signal. This would not be possible for other homonuclear alkali-metal pairs, and actually this two-photon excited channel should be taken into account in the model for OS. In the present case, such a two-photon excited channel is irrelevant. Therefore the idea would be to look for hyperfine-changing collisions (hcc) \cite{santos1999,young2000,marcassa2000,mudrich2004b,wang2020} leading to trap loss, especially in shallow traps. This phenomena is taken into account in an effective way in our model through the $P_S$ parameter. Such collisions have long been considered as a serious issue to achieve quantum degeneracy, and now as an additional possibility for controlling the appearance of new quantum phases in quantum gases \cite{wang2020}. Therefore, the OS would manifest by a reduction of losses from the trap, when one of the atoms is prepared in an excited level of the ground state hyperfine manifold. In the recent investigation of a $^6$Li-$^{41}$K ultracold mixture \cite{wang2020}, the final state of the hcc products is measured, and OS would then be detectable from such measurements.
 

\section{Conclusion}

Our calculations demonstrate the possibility to achieve OS of ultracold K-Cs collisions by using experimentally accessible parameters for the OS laser. One of the main difference with the homonuclear case, beside the range of the atom-atom interaction (varying at large distance as $R^{-6}$ instead of $R^{-3}$ for the homonuclear case) is that the OS efficiency does not depend on the polarization of the OS laser. As the behaviour of the relevant potential energy curves of all heteronuclear alkali-metal atom pairs are very similar (see for instance \cite{marinescu1999,beuc2006,beuc2016}), as well as comparable radiative lifetimes, it is likely that OS can be generalized to all of them.

If such an OS experiment would be implemented, the issue of the shielding of collisions between identical atoms with the mixture would still remain, \textit{i.e.} between potassium atoms in the present case. Assuming that Cs atoms would be created in their lowest hyperfine level, then the OS laser which is tuned to the blue of the potassium resonant transition would contribute to the shielding of K-K collisions too, providing that the polarization is chosen circular, as discussed in ref. \cite{weiner1999,suominen1995,napolitano1997}. These statements insure that as K-Cs collisions are shielded, so that three-body recombination (3BR) involving one Cs atom and two K atoms would be strongly hindered. If the fermionic potassium atom is chosen such 3BR will be even more hindered at low-temperature due to Pauli exclusion principle.

The present model already assumes a quite low temperature, which, in classical terms, determines the time spent by the system in the excited channel. As the temperature decreases, this excursion time decreases thus the influence of the spontaneous emission too. Moreover the collision energy becomes smaller and smaller compared to the chosen Rabi frequency, so that the light-induced avoided crossing turns more and more adiabatic, enforcing the shielding efficiency. Therefore we predict that the quantum degeneracy could be driven while keeping on the light-induced artificial interactions.

None of the numerous ongoing experiments devoted to ultracold binary mixtures of atomic species has investigated the possibility of OS, as researchers had different objectives, like the creation of ultracold molecules. However we note that in some experiments (e.g. Ref. \cite{burchianti2018} on double BEC of $^{41}$K and $^{87}$Rb), huge inelastic losses were reported, hindering to some extent the implementation of the dual-species BEC. Such a configuration would be a convenient test bed for OS, before trying it on molecular ensembles.

 \section{Acknowledgements}
 This work has been supported by the BLUESHIELD project (ANR-14-CE34-0006 from \textit{Agence Nationale de la Recherche}), by \textit{Investissements d'Avenir} LabEx PALM (ANR-10-LABX-0039-PALM), by \textit{R\'egion Ile-de-France} in the framework of DIM SIRTEQ, by the program Hubert Curien "BALATON" (Campus France Grant No.41919RK), and by the National Research Development and Innovation Office-NKFIH, 2018-2.1.13-T\'ET-FR-201800022. T. X. acknowledges the support from National Science Foundation of China under Grants No. 22103085. A. O. acknowledges partial support from National Research, Development and Innovation Office (Grant No. 2018-1.2.1-NKP-2018-00010 and K18, FK19 funding schemes with projects no. K 128621, FK 132989). X. X. acknowledges support from the Chinese Scholarship Council (Grant No. 201706240178). We are thankful to Dr Maxence Lepers for fruitful discussions.  Calculations have been achieved at the computing center M\'esoLUM of LUMAT (Research federation FR2764 of cNRS and \textit{Facult\'{e} des Sciences d'Orsay}).

\section{Appendix}

\begin{figure}[!t]\centering
\resizebox{0.45\textwidth}{!}{
  \includegraphics{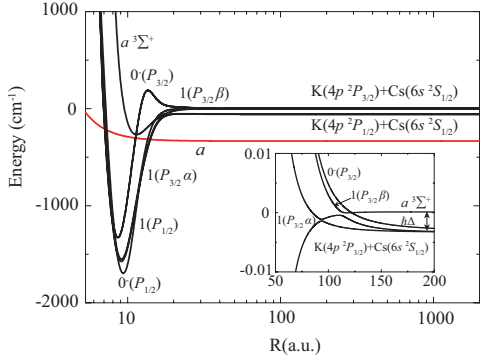}
} 
\caption{
The dressed adiabatic PECs in the space-fixed frame, setting $J = 0$ in the $a^3\Sigma^+$ entrance channel with $\bar{N}$ photons, coupled with a linearly-polarized field to $J' = 1$ in the excited states with $\bar{N}-1$ photons for a Rabi frequency $\omega = 10$~MHz. At infinity, the PECs are separated by the energy $\hbar \Delta$ with the detuning $\Delta = 200$~MHz. For simplification, only $f$-parity states are displayed, consistently with the selection rule for a $J\rightarrow J+1$ transition. The excited PECs are labeled with their dominant Hund's case $c$ character as SOC is included. The indexes $\alpha$ and $\beta$ distinguish the two 1($P_{3/2}$) states. The artificial channel (red curve) is noted as \emph{a}. The inset displays the avoided crossings responsible for the dynamics of optical shielding.
}
\label{fig:dressed_triplet}
\end{figure}

For completeness, we display in Figure \ref{fig:dressed_triplet} the PECs relevant when the entrance channel is the $a^3\Sigma^+$ state involving $f$ parity states. It is presented in the same way than Fig. \ref{fig:dressed} for the $X^1\Sigma^+$ entrance channel, and is thus helpful to better understand the discussion about the polarization dependence based on Figs. \ref{fig:DPECS10},  \ref{fig:DPECS200}. We note here the appearance of the $0^-$ excited molecular state which is composed of purely triplet Hund's case $a$ states, namely $(2)^3\Sigma^+$ and $(1)^3\Pi$. In contrast the $0^+$ states, which are of $e$ parity for $J=0$, do not contribute to the figure.

\begin{thebibliography}{73}%
\makeatletter
\providecommand \@ifxundefined [1]{%
 \@ifx{#1\undefined}
}%
\providecommand \@ifnum [1]{%
 \ifnum #1\expandafter \@firstoftwo
 \else \expandafter \@secondoftwo
 \fi
}%
\providecommand \@ifx [1]{%
 \ifx #1\expandafter \@firstoftwo
 \else \expandafter \@secondoftwo
 \fi
}%
\providecommand \natexlab [1]{#1}%
\providecommand \enquote  [1]{``#1''}%
\providecommand \bibnamefont  [1]{#1}%
\providecommand \bibfnamefont [1]{#1}%
\providecommand \citenamefont [1]{#1}%
\providecommand \href@noop [0]{\@secondoftwo}%
\providecommand \href [0]{\begingroup \@sanitize@url \@href}%
\providecommand \@href[1]{\@@startlink{#1}\@@href}%
\providecommand \@@href[1]{\endgroup#1\@@endlink}%
\providecommand \@sanitize@url [0]{\catcode `\\12\catcode `\$12\catcode
  `\&12\catcode `\#12\catcode `\^12\catcode `\_12\catcode `\%12\relax}%
\providecommand \@@startlink[1]{}%
\providecommand \@@endlink[0]{}%
\providecommand \url  [0]{\begingroup\@sanitize@url \@url }%
\providecommand \@url [1]{\endgroup\@href {#1}{\urlprefix }}%
\providecommand \urlprefix  [0]{URL }%
\providecommand \Eprint [0]{\href }%
\providecommand \doibase [0]{http://dx.doi.org/}%
\providecommand \selectlanguage [0]{\@gobble}%
\providecommand \bibinfo  [0]{\@secondoftwo}%
\providecommand \bibfield  [0]{\@secondoftwo}%
\providecommand \translation [1]{[#1]}%
\providecommand \BibitemOpen [0]{}%
\providecommand \bibitemStop [0]{}%
\providecommand \bibitemNoStop [0]{.\EOS\space}%
\providecommand \EOS [0]{\spacefactor3000\relax}%
\providecommand \BibitemShut  [1]{\csname bibitem#1\endcsname}%
\let\auto@bib@innerbib\@empty
\bibitem [{\citenamefont {Baranov}(2007)}]{baranov2008}%
  \BibitemOpen
  \bibfield  {author} {\bibinfo {author} {\bibfnamefont {M.~A.}\ \bibnamefont
  {Baranov}},\ }\href@noop {} {\bibfield  {journal} {\bibinfo  {journal} {Phys.
  Rep.}\ }\textbf {\bibinfo {volume} {464}},\ \bibinfo {pages} {71} (\bibinfo
  {year} {2007})}\BibitemShut {NoStop}%
\bibitem [{\citenamefont {Bloch}\ \emph {et~al.}(2008)\citenamefont {Bloch},
  \citenamefont {Dalibard},\ and\ \citenamefont {Zwerger}}]{bloch2008}%
  \BibitemOpen
  \bibfield  {author} {\bibinfo {author} {\bibfnamefont {I.}~\bibnamefont
  {Bloch}}, \bibinfo {author} {\bibfnamefont {J.}~\bibnamefont {Dalibard}}, \
  and\ \bibinfo {author} {\bibfnamefont {W.}~\bibnamefont {Zwerger}},\
  }\href@noop {} {\bibfield  {journal} {\bibinfo  {journal} {Rev. Mod. Phys.}\
  }\textbf {\bibinfo {volume} {80}},\ \bibinfo {pages} {885} (\bibinfo {year}
  {2008})}\BibitemShut {NoStop}%
\bibitem [{\citenamefont {Lahaye}\ \emph {et~al.}(2009)\citenamefont {Lahaye},
  \citenamefont {Menotti}, \citenamefont {Santos}, \citenamefont {Lewenstein},\
  and\ \citenamefont {Pfau}}]{lahaye2009}%
  \BibitemOpen
  \bibfield  {author} {\bibinfo {author} {\bibfnamefont {T.}~\bibnamefont
  {Lahaye}}, \bibinfo {author} {\bibfnamefont {C.}~\bibnamefont {Menotti}},
  \bibinfo {author} {\bibfnamefont {L.}~\bibnamefont {Santos}}, \bibinfo
  {author} {\bibfnamefont {M.}~\bibnamefont {Lewenstein}}, \ and\ \bibinfo
  {author} {\bibfnamefont {T.}~\bibnamefont {Pfau}},\ }\href@noop {} {\bibfield
   {journal} {\bibinfo  {journal} {Rep. Prog. Phys.}\ }\textbf {\bibinfo
  {volume} {72}},\ \bibinfo {pages} {126401} (\bibinfo {year}
  {2009})}\BibitemShut {NoStop}%
\bibitem [{\citenamefont {Bloch}\ \emph {et~al.}(2005)\citenamefont {Bloch},
  \citenamefont {Dalibard},\ and\ \citenamefont {Nascimb\`ene}}]{bloch2012}%
  \BibitemOpen
  \bibfield  {author} {\bibinfo {author} {\bibfnamefont {I.}~\bibnamefont
  {Bloch}}, \bibinfo {author} {\bibfnamefont {J.}~\bibnamefont {Dalibard}}, \
  and\ \bibinfo {author} {\bibfnamefont {S.}~\bibnamefont {Nascimb\`ene}},\
  }\href@noop {} {\bibfield  {journal} {\bibinfo  {journal} {Nature Phys.}\
  }\textbf {\bibinfo {volume} {38}},\ \bibinfo {pages} {S629} (\bibinfo {year}
  {2005})}\BibitemShut {NoStop}%
\bibitem [{\citenamefont {Doyle}\ \emph {et~al.}(2004)\citenamefont {Doyle},
  \citenamefont {Friedrich}, \citenamefont {Krems},\ and\ \citenamefont
  {Masnou-Seeuws}}]{doyle2004}%
  \BibitemOpen
  \bibfield  {author} {\bibinfo {author} {\bibfnamefont {J.}~\bibnamefont
  {Doyle}}, \bibinfo {author} {\bibfnamefont {B.}~\bibnamefont {Friedrich}},
  \bibinfo {author} {\bibfnamefont {R.}~\bibnamefont {Krems}}, \ and\ \bibinfo
  {author} {\bibfnamefont {F.}~\bibnamefont {Masnou-Seeuws}},\ }\href@noop {}
  {\bibfield  {journal} {\bibinfo  {journal} {Eur. Phys. J. D}\ }\textbf
  {\bibinfo {volume} {31}},\ \bibinfo {pages} {149} (\bibinfo {year}
  {2004})}\BibitemShut {NoStop}%
\bibitem [{\citenamefont {Carr}\ \emph {et~al.}(2009)\citenamefont {Carr},
  \citenamefont {DeMille}, \citenamefont {Krems},\ and\ \citenamefont
  {Ye}}]{carr2009b}%
  \BibitemOpen
  \bibfield  {author} {\bibinfo {author} {\bibfnamefont {L.~D.}\ \bibnamefont
  {Carr}}, \bibinfo {author} {\bibfnamefont {D.}~\bibnamefont {DeMille}},
  \bibinfo {author} {\bibfnamefont {R.~V.}\ \bibnamefont {Krems}}, \ and\
  \bibinfo {author} {\bibfnamefont {J.}~\bibnamefont {Ye}},\ }\href@noop {}
  {\bibfield  {journal} {\bibinfo  {journal} {New J. Phys.}\ }\textbf {\bibinfo
  {volume} {11}},\ \bibinfo {pages} {055049} (\bibinfo {year}
  {2009})}\BibitemShut {NoStop}%
\bibitem [{\citenamefont {Dulieu}\ and\ \citenamefont
  {Gabbanini}(2009)}]{dulieu2009}%
  \BibitemOpen
  \bibfield  {author} {\bibinfo {author} {\bibfnamefont {O.}~\bibnamefont
  {Dulieu}}\ and\ \bibinfo {author} {\bibfnamefont {C.}~\bibnamefont
  {Gabbanini}},\ }\href@noop {} {\bibfield  {journal} {\bibinfo  {journal}
  {Rep. Prog. Phys.}\ }\textbf {\bibinfo {volume} {72}},\ \bibinfo {pages}
  {086401} (\bibinfo {year} {2009})}\BibitemShut {NoStop}%
\bibitem [{\citenamefont {Qu{\'e}m{\'e}ner}\ and\ \citenamefont
  {Julienne}(2012)}]{quemener2012}%
  \BibitemOpen
  \bibfield  {author} {\bibinfo {author} {\bibfnamefont {G.}~\bibnamefont
  {Qu{\'e}m{\'e}ner}}\ and\ \bibinfo {author} {\bibfnamefont {P.~S.}\
  \bibnamefont {Julienne}},\ }\href@noop {} {\bibfield  {journal} {\bibinfo
  {journal} {Chem. Rev.}\ }\textbf {\bibinfo {volume} {112}},\ \bibinfo {pages}
  {4949} (\bibinfo {year} {2012})}\BibitemShut {NoStop}%
\bibitem [{\citenamefont {Moses}\ \emph {et~al.}(2017)\citenamefont {Moses},
  \citenamefont {Covey}, \citenamefont {Miecnikowski}, \citenamefont {Jin},\
  and\ \citenamefont {Ye}}]{moses2017}%
  \BibitemOpen
  \bibfield  {author} {\bibinfo {author} {\bibfnamefont {S.~A.}\ \bibnamefont
  {Moses}}, \bibinfo {author} {\bibfnamefont {J.~P.}\ \bibnamefont {Covey}},
  \bibinfo {author} {\bibfnamefont {M.~T.}\ \bibnamefont {Miecnikowski}},
  \bibinfo {author} {\bibfnamefont {D.~S.}\ \bibnamefont {Jin}}, \ and\
  \bibinfo {author} {\bibfnamefont {J.}~\bibnamefont {Ye}},\ }\href@noop {}
  {\bibfield  {journal} {\bibinfo  {journal} {Nature Phys.}\ }\textbf {\bibinfo
  {volume} {13}},\ \bibinfo {pages} {13} (\bibinfo {year} {2017})}\BibitemShut
  {NoStop}%
\bibitem [{\citenamefont {Bohn}\ \emph {et~al.}(2017)\citenamefont {Bohn},
  \citenamefont {Rey},\ and\ \citenamefont {Ye}}]{bohn2017}%
  \BibitemOpen
  \bibfield  {author} {\bibinfo {author} {\bibfnamefont {J.~L.}\ \bibnamefont
  {Bohn}}, \bibinfo {author} {\bibfnamefont {A.~M.}\ \bibnamefont {Rey}}, \
  and\ \bibinfo {author} {\bibfnamefont {J.}~\bibnamefont {Ye}},\ }\href@noop
  {} {\bibfield  {journal} {\bibinfo  {journal} {Science}\ }\textbf {\bibinfo
  {volume} {357}},\ \bibinfo {pages} {1002} (\bibinfo {year}
  {2017})}\BibitemShut {NoStop}%
\bibitem [{\citenamefont {Ospelkaus}\ \emph {et~al.}(2010)\citenamefont
  {Ospelkaus}, \citenamefont {Ni}, \citenamefont {Wang}, \citenamefont
  {de~Miranda}, \citenamefont {Neyenhuis}, \citenamefont {Qu\'em\'ener},
  \citenamefont {Julienne}, \citenamefont {Bohn}, \citenamefont {Jin},\ and\
  \citenamefont {Ye}}]{ospelkaus2010a}%
  \BibitemOpen
  \bibfield  {author} {\bibinfo {author} {\bibfnamefont {S.}~\bibnamefont
  {Ospelkaus}}, \bibinfo {author} {\bibfnamefont {K.-K.}\ \bibnamefont {Ni}},
  \bibinfo {author} {\bibfnamefont {D.}~\bibnamefont {Wang}}, \bibinfo {author}
  {\bibfnamefont {M.~H.~G.}\ \bibnamefont {de~Miranda}}, \bibinfo {author}
  {\bibfnamefont {B.}~\bibnamefont {Neyenhuis}}, \bibinfo {author}
  {\bibfnamefont {G.}~\bibnamefont {Qu\'em\'ener}}, \bibinfo {author}
  {\bibfnamefont {P.~S.}\ \bibnamefont {Julienne}}, \bibinfo {author}
  {\bibfnamefont {J.}~\bibnamefont {Bohn}}, \bibinfo {author} {\bibfnamefont
  {D.~S.}\ \bibnamefont {Jin}}, \ and\ \bibinfo {author} {\bibfnamefont
  {J.}~\bibnamefont {Ye}},\ }\href@noop {} {\bibfield  {journal} {\bibinfo
  {journal} {Science}\ }\textbf {\bibinfo {volume} {327}},\ \bibinfo {pages}
  {853} (\bibinfo {year} {2010})}\BibitemShut {NoStop}%
\bibitem [{\citenamefont {Ni}\ \emph {et~al.}(2010)\citenamefont {Ni},
  \citenamefont {Ospelkaus}, \citenamefont {Wang}, \citenamefont
  {Qu\'em\'ener}, \citenamefont {Neyenhuis}, \citenamefont {de~Miranda},
  \citenamefont {Bohn}, \citenamefont {Ye},\ and\ \citenamefont
  {Jin}}]{ni2010}%
  \BibitemOpen
  \bibfield  {author} {\bibinfo {author} {\bibfnamefont {K.-K.}\ \bibnamefont
  {Ni}}, \bibinfo {author} {\bibfnamefont {S.}~\bibnamefont {Ospelkaus}},
  \bibinfo {author} {\bibfnamefont {D.}~\bibnamefont {Wang}}, \bibinfo {author}
  {\bibfnamefont {G.}~\bibnamefont {Qu\'em\'ener}}, \bibinfo {author}
  {\bibfnamefont {B.}~\bibnamefont {Neyenhuis}}, \bibinfo {author}
  {\bibfnamefont {M.~H.~G.}\ \bibnamefont {de~Miranda}}, \bibinfo {author}
  {\bibfnamefont {J.~L.}\ \bibnamefont {Bohn}}, \bibinfo {author}
  {\bibfnamefont {J.}~\bibnamefont {Ye}}, \ and\ \bibinfo {author}
  {\bibfnamefont {D.~S.}\ \bibnamefont {Jin}},\ }\href@noop {} {\bibfield
  {journal} {\bibinfo  {journal} {Nature}\ }\textbf {\bibinfo {volume} {464}},\
  \bibinfo {pages} {1324} (\bibinfo {year} {2010})}\BibitemShut {NoStop}%
\bibitem [{\citenamefont {Takekoshi}\ \emph {et~al.}(2014)\citenamefont
  {Takekoshi}, \citenamefont {Reichs\"ollner}, \citenamefont {Schindewolf},
  \citenamefont {Hutson}, \citenamefont {Sueur}, \citenamefont {Dulieu},
  \citenamefont {Ferlaino}, \citenamefont {Grimm},\ and\ \citenamefont
  {N\"agerl}}]{takekoshi2014}%
  \BibitemOpen
  \bibfield  {author} {\bibinfo {author} {\bibfnamefont {T.}~\bibnamefont
  {Takekoshi}}, \bibinfo {author} {\bibfnamefont {L.}~\bibnamefont
  {Reichs\"ollner}}, \bibinfo {author} {\bibfnamefont {A.}~\bibnamefont
  {Schindewolf}}, \bibinfo {author} {\bibfnamefont {J.~M.}\ \bibnamefont
  {Hutson}}, \bibinfo {author} {\bibfnamefont {C.~R.~L.}\ \bibnamefont
  {Sueur}}, \bibinfo {author} {\bibfnamefont {O.}~\bibnamefont {Dulieu}},
  \bibinfo {author} {\bibfnamefont {F.}~\bibnamefont {Ferlaino}}, \bibinfo
  {author} {\bibfnamefont {R.}~\bibnamefont {Grimm}}, \ and\ \bibinfo {author}
  {\bibfnamefont {H.-C.}\ \bibnamefont {N\"agerl}},\ }\href@noop {} {\bibfield
  {journal} {\bibinfo  {journal} {Phys. Rev. Lett.}\ }\textbf {\bibinfo
  {volume} {113}},\ \bibinfo {pages} {205301} (\bibinfo {year}
  {2014})}\BibitemShut {NoStop}%
\bibitem [{\citenamefont {Molony}\ \emph {et~al.}(2014)\citenamefont {Molony},
  \citenamefont {Gregory}, \citenamefont {Ji}, \citenamefont {Lu},
  \citenamefont {K\"oppinger}, \citenamefont {Le~Sueur}, \citenamefont
  {Blackley}, \citenamefont {Hutson},\ and\ \citenamefont
  {Cornish}}]{molony2014}%
  \BibitemOpen
  \bibfield  {author} {\bibinfo {author} {\bibfnamefont {P.~K.}\ \bibnamefont
  {Molony}}, \bibinfo {author} {\bibfnamefont {P.~D.}\ \bibnamefont {Gregory}},
  \bibinfo {author} {\bibfnamefont {Z.}~\bibnamefont {Ji}}, \bibinfo {author}
  {\bibfnamefont {B.}~\bibnamefont {Lu}}, \bibinfo {author} {\bibfnamefont
  {M.~P.}\ \bibnamefont {K\"oppinger}}, \bibinfo {author} {\bibfnamefont
  {C.~R.}\ \bibnamefont {Le~Sueur}}, \bibinfo {author} {\bibfnamefont {C.~L.}\
  \bibnamefont {Blackley}}, \bibinfo {author} {\bibfnamefont {J.~M.}\
  \bibnamefont {Hutson}}, \ and\ \bibinfo {author} {\bibfnamefont {S.~L.}\
  \bibnamefont {Cornish}},\ }\href@noop {} {\bibfield  {journal} {\bibinfo
  {journal} {Phys. Rev. Lett.}\ }\textbf {\bibinfo {volume} {113}},\ \bibinfo
  {pages} {255301} (\bibinfo {year} {2014})}\BibitemShut {NoStop}%
\bibitem [{\citenamefont {Guo}\ \emph {et~al.}(2016)\citenamefont {Guo},
  \citenamefont {Zhu}, \citenamefont {Lu}, \citenamefont {Ye}, \citenamefont
  {Wang}, \citenamefont {Vexiau}, \citenamefont {Bouloufa-Maafa}, \citenamefont
  {Qu\'em\'ener}, \citenamefont {Dulieu},\ and\ \citenamefont
  {Wang}}]{guo2016}%
  \BibitemOpen
  \bibfield  {author} {\bibinfo {author} {\bibfnamefont {M.}~\bibnamefont
  {Guo}}, \bibinfo {author} {\bibfnamefont {B.}~\bibnamefont {Zhu}}, \bibinfo
  {author} {\bibfnamefont {B.}~\bibnamefont {Lu}}, \bibinfo {author}
  {\bibfnamefont {X.}~\bibnamefont {Ye}}, \bibinfo {author} {\bibfnamefont
  {F.}~\bibnamefont {Wang}}, \bibinfo {author} {\bibfnamefont {R.}~\bibnamefont
  {Vexiau}}, \bibinfo {author} {\bibfnamefont {N.}~\bibnamefont
  {Bouloufa-Maafa}}, \bibinfo {author} {\bibfnamefont {G.}~\bibnamefont
  {Qu\'em\'ener}}, \bibinfo {author} {\bibfnamefont {O.}~\bibnamefont
  {Dulieu}}, \ and\ \bibinfo {author} {\bibfnamefont {D.}~\bibnamefont
  {Wang}},\ }\href@noop {} {\bibfield  {journal} {\bibinfo  {journal} {Phys.
  Rev. Lett.}\ }\textbf {\bibinfo {volume} {116}},\ \bibinfo {pages} {205303}
  (\bibinfo {year} {2016})}\BibitemShut {NoStop}%
\bibitem [{\citenamefont {Gregory}\ \emph {et~al.}(2019)\citenamefont
  {Gregory}, \citenamefont {Frye}, \citenamefont {Blackmore}, \citenamefont
  {Bridge}, \citenamefont {Sawant}, \citenamefont {Hutson}, ,\ and\
  \citenamefont {Cornish}}]{gregory2019}%
  \BibitemOpen
  \bibfield  {author} {\bibinfo {author} {\bibfnamefont {P.~D.}\ \bibnamefont
  {Gregory}}, \bibinfo {author} {\bibfnamefont {M.~D.}\ \bibnamefont {Frye}},
  \bibinfo {author} {\bibfnamefont {J.~A.}\ \bibnamefont {Blackmore}}, \bibinfo
  {author} {\bibfnamefont {E.~M.}\ \bibnamefont {Bridge}}, \bibinfo {author}
  {\bibfnamefont {R.}~\bibnamefont {Sawant}}, \bibinfo {author} {\bibfnamefont
  {J.~M.}\ \bibnamefont {Hutson}}, , \ and\ \bibinfo {author} {\bibfnamefont
  {S.~L.}\ \bibnamefont {Cornish}},\ }\href@noop {} {\bibfield  {journal}
  {\bibinfo  {journal} {Nature Comm.}\ }\textbf {\bibinfo {volume} {10}},\
  \bibinfo {pages} {3104} (\bibinfo {year} {2019})}\BibitemShut {NoStop}%
\bibitem [{\citenamefont {Mayle}\ \emph {et~al.}(2013)\citenamefont {Mayle},
  \citenamefont {Qu\'em\'ener}, \citenamefont {Ruzic},\ and\ \citenamefont
  {Bohn}}]{mayle2013}%
  \BibitemOpen
  \bibfield  {author} {\bibinfo {author} {\bibfnamefont {M.}~\bibnamefont
  {Mayle}}, \bibinfo {author} {\bibfnamefont {G.}~\bibnamefont {Qu\'em\'ener}},
  \bibinfo {author} {\bibfnamefont {B.~P.}\ \bibnamefont {Ruzic}}, \ and\
  \bibinfo {author} {\bibfnamefont {J.~L.}\ \bibnamefont {Bohn}},\ }\href@noop
  {} {\bibfield  {journal} {\bibinfo  {journal} {Phys. Rev. A}\ }\textbf
  {\bibinfo {volume} {87}},\ \bibinfo {pages} {012709} (\bibinfo {year}
  {2013})}\BibitemShut {NoStop}%
\bibitem [{\citenamefont {Croft}\ and\ \citenamefont {Bohn}(2014)}]{croft2014}%
  \BibitemOpen
  \bibfield  {author} {\bibinfo {author} {\bibfnamefont {J.~F.~E.}\
  \bibnamefont {Croft}}\ and\ \bibinfo {author} {\bibfnamefont {J.~L.}\
  \bibnamefont {Bohn}},\ }\href@noop {} {\bibfield  {journal} {\bibinfo
  {journal} {Phys. Rev. A}\ }\textbf {\bibinfo {volume} {89}},\ \bibinfo
  {pages} {012714} (\bibinfo {year} {2014})}\BibitemShut {NoStop}%
\bibitem [{\citenamefont {Christianen}\ \emph
  {et~al.}(2019{\natexlab{a}})\citenamefont {Christianen}, \citenamefont
  {Zwierlein}, \citenamefont {Groenenboom},\ and\ \citenamefont
  {Karman}}]{christianen2019a}%
  \BibitemOpen
  \bibfield  {author} {\bibinfo {author} {\bibfnamefont {A.}~\bibnamefont
  {Christianen}}, \bibinfo {author} {\bibfnamefont {M.~W.}\ \bibnamefont
  {Zwierlein}}, \bibinfo {author} {\bibfnamefont {G.~C.}\ \bibnamefont
  {Groenenboom}}, \ and\ \bibinfo {author} {\bibfnamefont {T.}~\bibnamefont
  {Karman}},\ }\href@noop {} {\bibfield  {journal} {\bibinfo  {journal} {Phys.
  Rev. Lett.}\ }\textbf {\bibinfo {volume} {123}},\ \bibinfo {pages} {123402}
  (\bibinfo {year} {2019}{\natexlab{a}})}\BibitemShut {NoStop}%
\bibitem [{\citenamefont {Christianen}\ \emph
  {et~al.}(2019{\natexlab{b}})\citenamefont {Christianen}, \citenamefont
  {Karman},\ and\ \citenamefont {Groenenboom}}]{christianen2019b}%
  \BibitemOpen
  \bibfield  {author} {\bibinfo {author} {\bibfnamefont {A.}~\bibnamefont
  {Christianen}}, \bibinfo {author} {\bibfnamefont {T.}~\bibnamefont {Karman}},
  \ and\ \bibinfo {author} {\bibfnamefont {G.~C.}\ \bibnamefont
  {Groenenboom}},\ }\href@noop {} {\bibfield  {journal} {\bibinfo  {journal}
  {Phys. Rev. A}\ }\textbf {\bibinfo {volume} {100}},\ \bibinfo {pages}
  {032708} (\bibinfo {year} {2019}{\natexlab{b}})}\BibitemShut {NoStop}%
\bibitem [{\citenamefont {Jachymski}\ \emph {et~al.}(2021)\citenamefont
  {Jachymski}, \citenamefont {Gronowski},\ and\ \citenamefont
  {Tomza}}]{jachymski2021}%
  \BibitemOpen
  \bibfield  {author} {\bibinfo {author} {\bibfnamefont {K.}~\bibnamefont
  {Jachymski}}, \bibinfo {author} {\bibfnamefont {M.}~\bibnamefont
  {Gronowski}}, \ and\ \bibinfo {author} {\bibfnamefont {M.}~\bibnamefont
  {Tomza}},\ }\href@noop {} {\enquote {\bibinfo {title} {Collisional losses of
  ultracold molecules due to intermediate complex formation},}\ } (\bibinfo
  {year} {2021}),\ \Eprint {http://arxiv.org/abs/2110.07501} {arXiv:2110.07501
  [cond-mat.quant-gas]} \BibitemShut {NoStop}%
\bibitem [{\citenamefont {Gregory}\ \emph {et~al.}(2020)\citenamefont
  {Gregory}, \citenamefont {Blackmore}, \citenamefont {Bromley},\ and\
  \citenamefont {Cornish}}]{gregory2020}%
  \BibitemOpen
  \bibfield  {author} {\bibinfo {author} {\bibfnamefont {P.~D.}\ \bibnamefont
  {Gregory}}, \bibinfo {author} {\bibfnamefont {J.~A.}\ \bibnamefont
  {Blackmore}}, \bibinfo {author} {\bibfnamefont {S.~L.}\ \bibnamefont
  {Bromley}}, \ and\ \bibinfo {author} {\bibfnamefont {S.~L.}\ \bibnamefont
  {Cornish}},\ }\href@noop {} {\bibfield  {journal} {\bibinfo  {journal} {Phys.
  Rev. Lett.}\ }\textbf {\bibinfo {volume} {124}},\ \bibinfo {pages} {163402}
  (\bibinfo {year} {2020})}\BibitemShut {NoStop}%
\bibitem [{\citenamefont {Liu}\ \emph {et~al.}(2020)\citenamefont {Liu},
  \citenamefont {Hu}, \citenamefont {Nichols}, \citenamefont {Grimes},
  \citenamefont {Karman}, \citenamefont {Guo},\ and\ \citenamefont
  {Ni}}]{liu2020}%
  \BibitemOpen
  \bibfield  {author} {\bibinfo {author} {\bibfnamefont {Y.}~\bibnamefont
  {Liu}}, \bibinfo {author} {\bibfnamefont {M.-G.}\ \bibnamefont {Hu}},
  \bibinfo {author} {\bibfnamefont {M.~A.}\ \bibnamefont {Nichols}}, \bibinfo
  {author} {\bibfnamefont {D.~D.}\ \bibnamefont {Grimes}}, \bibinfo {author}
  {\bibfnamefont {T.}~\bibnamefont {Karman}}, \bibinfo {author} {\bibfnamefont
  {H.}~\bibnamefont {Guo}}, \ and\ \bibinfo {author} {\bibfnamefont {K.-K.}\
  \bibnamefont {Ni}},\ }\href@noop {} {\bibfield  {journal} {\bibinfo
  {journal} {Nature Phys.}\ ,\ \bibinfo {pages}
  {https://doi.org/10.1038/s41567}} (\bibinfo {year} {2020})}\BibitemShut
  {NoStop}%
\bibitem [{\citenamefont {Bause}\ \emph {et~al.}(2021)\citenamefont {Bause},
  \citenamefont {Schindewolf}, \citenamefont {Tao}, \citenamefont {Duda},
  \citenamefont {Chen}, \citenamefont {Qu\'em\'ener}, \citenamefont {Karman},
  \citenamefont {Christianen}, \citenamefont {Bloch},\ and\ \citenamefont
  {Luo}}]{bause2021}%
  \BibitemOpen
  \bibfield  {author} {\bibinfo {author} {\bibfnamefont {R.}~\bibnamefont
  {Bause}}, \bibinfo {author} {\bibfnamefont {A.}~\bibnamefont {Schindewolf}},
  \bibinfo {author} {\bibfnamefont {R.}~\bibnamefont {Tao}}, \bibinfo {author}
  {\bibfnamefont {M.}~\bibnamefont {Duda}}, \bibinfo {author} {\bibfnamefont
  {X.-Y.}\ \bibnamefont {Chen}}, \bibinfo {author} {\bibfnamefont
  {G.}~\bibnamefont {Qu\'em\'ener}}, \bibinfo {author} {\bibfnamefont
  {T.}~\bibnamefont {Karman}}, \bibinfo {author} {\bibfnamefont
  {A.}~\bibnamefont {Christianen}}, \bibinfo {author} {\bibfnamefont
  {I.}~\bibnamefont {Bloch}}, \ and\ \bibinfo {author} {\bibfnamefont {X.-Y.}\
  \bibnamefont {Luo}},\ }\href@noop {} {\bibfield  {journal} {\bibinfo
  {journal} {Phys. Rev. Research}\ }\textbf {\bibinfo {volume} {3}},\ \bibinfo
  {pages} {033013} (\bibinfo {year} {2021})}\BibitemShut {NoStop}%
\bibitem [{\citenamefont {Gersema}\ \emph {et~al.}(2021)\citenamefont
  {Gersema}, \citenamefont {Voges}, \citenamefont {Meyer~zum Alten~Borgloh},
  \citenamefont {Koch}, \citenamefont {Hartmann}, \citenamefont {Zenesini},
  \citenamefont {Ospelkaus}, \citenamefont {Lin}, \citenamefont {He},\ and\
  \citenamefont {Wang}}]{gersema2021}%
  \BibitemOpen
  \bibfield  {author} {\bibinfo {author} {\bibfnamefont {P.}~\bibnamefont
  {Gersema}}, \bibinfo {author} {\bibfnamefont {K.~K.}\ \bibnamefont {Voges}},
  \bibinfo {author} {\bibfnamefont {M.}~\bibnamefont {Meyer~zum
  Alten~Borgloh}}, \bibinfo {author} {\bibfnamefont {L.}~\bibnamefont {Koch}},
  \bibinfo {author} {\bibfnamefont {T.}~\bibnamefont {Hartmann}}, \bibinfo
  {author} {\bibfnamefont {A.}~\bibnamefont {Zenesini}}, \bibinfo {author}
  {\bibfnamefont {S.}~\bibnamefont {Ospelkaus}}, \bibinfo {author}
  {\bibfnamefont {J.}~\bibnamefont {Lin}}, \bibinfo {author} {\bibfnamefont
  {J.}~\bibnamefont {He}}, \ and\ \bibinfo {author} {\bibfnamefont
  {D.}~\bibnamefont {Wang}},\ }\href@noop {} {\bibfield  {journal} {\bibinfo
  {journal} {Phys. Rev. Lett.}\ }\textbf {\bibinfo {volume} {127}},\ \bibinfo
  {pages} {163401} (\bibinfo {year} {2021})}\BibitemShut {NoStop}%
\bibitem [{\citenamefont {Avdeenkov}\ \emph {et~al.}(2006)\citenamefont
  {Avdeenkov}, \citenamefont {Kajita},\ and\ \citenamefont
  {Bohn}}]{avdeenkov2006a}%
  \BibitemOpen
  \bibfield  {author} {\bibinfo {author} {\bibfnamefont {A.~V.}\ \bibnamefont
  {Avdeenkov}}, \bibinfo {author} {\bibfnamefont {M.}~\bibnamefont {Kajita}}, \
  and\ \bibinfo {author} {\bibfnamefont {J.~L.}\ \bibnamefont {Bohn}},\
  }\href@noop {} {\bibfield  {journal} {\bibinfo  {journal} {Phys. Rev. A}\
  }\textbf {\bibinfo {volume} {73}},\ \bibinfo {pages} {022707} (\bibinfo
  {year} {2006})}\BibitemShut {NoStop}%
\bibitem [{\citenamefont {Qu{\'e}m{\'e}ner}\ and\ \citenamefont
  {Bohn}(2010)}]{quemener2010b}%
  \BibitemOpen
  \bibfield  {author} {\bibinfo {author} {\bibfnamefont {G.}~\bibnamefont
  {Qu{\'e}m{\'e}ner}}\ and\ \bibinfo {author} {\bibfnamefont {J.}~\bibnamefont
  {Bohn}},\ }\href@noop {} {\bibfield  {journal} {\bibinfo  {journal} {Phys.
  Rev. A}\ }\textbf {\bibinfo {volume} {81}},\ \bibinfo {pages} {060701}
  (\bibinfo {year} {2010})}\BibitemShut {NoStop}%
\bibitem [{\citenamefont {Wang}\ and\ \citenamefont
  {Qu{\'{e}}m{\'{e}}ner}(2015)}]{wang2015}%
  \BibitemOpen
  \bibfield  {author} {\bibinfo {author} {\bibfnamefont {G.}~\bibnamefont
  {Wang}}\ and\ \bibinfo {author} {\bibfnamefont {G.}~\bibnamefont
  {Qu{\'{e}}m{\'{e}}ner}},\ }\href@noop {} {\bibfield  {journal} {\bibinfo
  {journal} {New J. Phys.}\ }\textbf {\bibinfo {volume} {17}},\ \bibinfo
  {pages} {035015} (\bibinfo {year} {2015})}\BibitemShut {NoStop}%
\bibitem [{\citenamefont {Qu\'em\'ener}\ and\ \citenamefont
  {Bohn}(2016)}]{quemener2016}%
  \BibitemOpen
  \bibfield  {author} {\bibinfo {author} {\bibfnamefont {G.}~\bibnamefont
  {Qu\'em\'ener}}\ and\ \bibinfo {author} {\bibfnamefont {J.~L.}\ \bibnamefont
  {Bohn}},\ }\href@noop {} {\bibfield  {journal} {\bibinfo  {journal} {Phys.
  Rev. A}\ }\textbf {\bibinfo {volume} {93}},\ \bibinfo {pages} {012704}
  (\bibinfo {year} {2016})}\BibitemShut {NoStop}%
\bibitem [{\citenamefont {Gonz\'alez-Mart\'{\i}nez}\ \emph
  {et~al.}(2017)\citenamefont {Gonz\'alez-Mart\'{\i}nez}, \citenamefont
  {Bohn},\ and\ \citenamefont {Qu\'em\'ener}}]{gonzalez-martinez2017}%
  \BibitemOpen
  \bibfield  {author} {\bibinfo {author} {\bibfnamefont {M.~L.}\ \bibnamefont
  {Gonz\'alez-Mart\'{\i}nez}}, \bibinfo {author} {\bibfnamefont {J.~L.}\
  \bibnamefont {Bohn}}, \ and\ \bibinfo {author} {\bibfnamefont
  {G.}~\bibnamefont {Qu\'em\'ener}},\ }\href@noop {} {\bibfield  {journal}
  {\bibinfo  {journal} {Phys. Rev. A}\ }\textbf {\bibinfo {volume} {96}},\
  \bibinfo {pages} {032718} (\bibinfo {year} {2017})}\BibitemShut {NoStop}%
\bibitem [{\citenamefont {Karman}\ and\ \citenamefont
  {Hutson}(2018)}]{karman2018}%
  \BibitemOpen
  \bibfield  {author} {\bibinfo {author} {\bibfnamefont {T.}~\bibnamefont
  {Karman}}\ and\ \bibinfo {author} {\bibfnamefont {J.~M.}\ \bibnamefont
  {Hutson}},\ }\href@noop {} {\bibfield  {journal} {\bibinfo  {journal} {Phys.
  Rev. Lett.}\ }\textbf {\bibinfo {volume} {121}},\ \bibinfo {pages} {163401}
  (\bibinfo {year} {2018})}\BibitemShut {NoStop}%
\bibitem [{\citenamefont {Lassabli\`ere}\ and\ \citenamefont
  {Qu\'em\'ener}(2018)}]{lassabliere2018}%
  \BibitemOpen
  \bibfield  {author} {\bibinfo {author} {\bibfnamefont {L.}~\bibnamefont
  {Lassabli\`ere}}\ and\ \bibinfo {author} {\bibfnamefont {G.}~\bibnamefont
  {Qu\'em\'ener}},\ }\href@noop {} {\bibfield  {journal} {\bibinfo  {journal}
  {Phys. Rev. Lett.}\ }\textbf {\bibinfo {volume} {121}},\ \bibinfo {pages}
  {163402} (\bibinfo {year} {2018})}\BibitemShut {NoStop}%
\bibitem [{\citenamefont {Karman}(2020)}]{karman2020}%
  \BibitemOpen
  \bibfield  {author} {\bibinfo {author} {\bibfnamefont {T.}~\bibnamefont
  {Karman}},\ }\href@noop {} {\bibfield  {journal} {\bibinfo  {journal} {Phys.
  Rev. A}\ }\textbf {\bibinfo {volume} {101}},\ \bibinfo {pages} {042702}
  (\bibinfo {year} {2020})}\BibitemShut {NoStop}%
\bibitem [{\citenamefont {Matsuda}\ \emph {et~al.}(2020)\citenamefont
  {Matsuda}, \citenamefont {Marco}, \citenamefont {Li}, \citenamefont {Tobias},
  \citenamefont {Valtolina}, \citenamefont {Qu\'em\'ener},\ and\ \citenamefont
  {Ye}}]{matsuda2020}%
  \BibitemOpen
  \bibfield  {author} {\bibinfo {author} {\bibfnamefont {K.}~\bibnamefont
  {Matsuda}}, \bibinfo {author} {\bibfnamefont {L.~D.}\ \bibnamefont {Marco}},
  \bibinfo {author} {\bibfnamefont {J.-R.}\ \bibnamefont {Li}}, \bibinfo
  {author} {\bibfnamefont {W.~G.}\ \bibnamefont {Tobias}}, \bibinfo {author}
  {\bibfnamefont {G.}~\bibnamefont {Valtolina}}, \bibinfo {author}
  {\bibfnamefont {G.}~\bibnamefont {Qu\'em\'ener}}, \ and\ \bibinfo {author}
  {\bibfnamefont {J.}~\bibnamefont {Ye}},\ }\href@noop {} {\bibfield  {journal}
  {\bibinfo  {journal} {Science}\ }\textbf {\bibinfo {volume} {370}},\ \bibinfo
  {pages} {1324} (\bibinfo {year} {2020})}\BibitemShut {NoStop}%
\bibitem [{\citenamefont {Li}\ \emph {et~al.}(2021)\citenamefont {Li},
  \citenamefont {Tobias}, \citenamefont {Matsuda}, \citenamefont {Miller},
  \citenamefont {Valtolina}, \citenamefont {Marco}, \citenamefont {Wang},
  \citenamefont {Lassabli\`ere}, \citenamefont {Qu\'em\'ener}, \citenamefont
  {Bohn},\ and\ \citenamefont {Ye}}]{li2021}%
  \BibitemOpen
  \bibfield  {author} {\bibinfo {author} {\bibfnamefont {J.-R.}\ \bibnamefont
  {Li}}, \bibinfo {author} {\bibfnamefont {W.~G.}\ \bibnamefont {Tobias}},
  \bibinfo {author} {\bibfnamefont {K.}~\bibnamefont {Matsuda}}, \bibinfo
  {author} {\bibfnamefont {C.}~\bibnamefont {Miller}}, \bibinfo {author}
  {\bibfnamefont {G.}~\bibnamefont {Valtolina}}, \bibinfo {author}
  {\bibfnamefont {L.~D.}\ \bibnamefont {Marco}}, \bibinfo {author}
  {\bibfnamefont {R.~R.}\ \bibnamefont {Wang}}, \bibinfo {author}
  {\bibfnamefont {L.}~\bibnamefont {Lassabli\`ere}}, \bibinfo {author}
  {\bibfnamefont {G.}~\bibnamefont {Qu\'em\'ener}}, \bibinfo {author}
  {\bibfnamefont {J.~L.}\ \bibnamefont {Bohn}}, \ and\ \bibinfo {author}
  {\bibfnamefont {J.}~\bibnamefont {Ye}},\ }\href@noop {} {\bibfield  {journal}
  {\bibinfo  {journal} {Nature Phys.}\ }\textbf {\bibinfo {volume} {17}},\
  \bibinfo {pages} {1144} (\bibinfo {year} {2021})}\BibitemShut {NoStop}%
\bibitem [{\citenamefont {Anderegg}\ \emph {et~al.}(2021)\citenamefont
  {Anderegg}, \citenamefont {Burchesky}, \citenamefont {Bao}, \citenamefont
  {Yu}, \citenamefont {Karman}, \citenamefont {Chae}, \citenamefont {Ni},
  \citenamefont {Ketterle},\ and\ \citenamefont {Doyle}}]{anderegg2021}%
  \BibitemOpen
  \bibfield  {author} {\bibinfo {author} {\bibfnamefont {L.}~\bibnamefont
  {Anderegg}}, \bibinfo {author} {\bibfnamefont {S.}~\bibnamefont {Burchesky}},
  \bibinfo {author} {\bibfnamefont {Y.}~\bibnamefont {Bao}}, \bibinfo {author}
  {\bibfnamefont {S.~S.}\ \bibnamefont {Yu}}, \bibinfo {author} {\bibfnamefont
  {T.}~\bibnamefont {Karman}}, \bibinfo {author} {\bibfnamefont
  {E.}~\bibnamefont {Chae}}, \bibinfo {author} {\bibfnamefont {K.-K.}\
  \bibnamefont {Ni}}, \bibinfo {author} {\bibfnamefont {W.}~\bibnamefont
  {Ketterle}}, \ and\ \bibinfo {author} {\bibfnamefont {J.~M.}\ \bibnamefont
  {Doyle}},\ }\href@noop {} {\bibfield  {journal} {\bibinfo  {journal}
  {Science}\ }\textbf {\bibinfo {volume} {373}},\ \bibinfo {pages} {779}
  (\bibinfo {year} {2021})}\BibitemShut {NoStop}%
\bibitem [{\citenamefont {Xie}\ \emph {et~al.}(2020)\citenamefont {Xie},
  \citenamefont {Lepers}, \citenamefont {Vexiau}, \citenamefont {Orb\'{a}n},
  \citenamefont {Dulieu},\ and\ \citenamefont {Bouloufa-Maafa}}]{xie2020}%
  \BibitemOpen
  \bibfield  {author} {\bibinfo {author} {\bibfnamefont {T.}~\bibnamefont
  {Xie}}, \bibinfo {author} {\bibfnamefont {M.}~\bibnamefont {Lepers}},
  \bibinfo {author} {\bibfnamefont {R.}~\bibnamefont {Vexiau}}, \bibinfo
  {author} {\bibfnamefont {A.}~\bibnamefont {Orb\'{a}n}}, \bibinfo {author}
  {\bibfnamefont {O.}~\bibnamefont {Dulieu}}, \ and\ \bibinfo {author}
  {\bibfnamefont {N.}~\bibnamefont {Bouloufa-Maafa}},\ }\href@noop {}
  {\bibfield  {journal} {\bibinfo  {journal} {Phys. Rev. Lett.}\ }\textbf
  {\bibinfo {volume} {125}},\ \bibinfo {pages} {153202} (\bibinfo {year}
  {2020})}\BibitemShut {NoStop}%
\bibitem [{\citenamefont {Zilio}\ \emph {et~al.}(1996)\citenamefont {Zilio},
  \citenamefont {Marcassa}, \citenamefont {Muniz}, \citenamefont {Horowicz},
  \citenamefont {Bagnato}, \citenamefont {Napolitano}, \citenamefont {Weiner},\
  and\ \citenamefont {Julienne}}]{zilio1996}%
  \BibitemOpen
  \bibfield  {author} {\bibinfo {author} {\bibfnamefont {S.}~\bibnamefont
  {Zilio}}, \bibinfo {author} {\bibfnamefont {L.}~\bibnamefont {Marcassa}},
  \bibinfo {author} {\bibfnamefont {S.}~\bibnamefont {Muniz}}, \bibinfo
  {author} {\bibfnamefont {R.}~\bibnamefont {Horowicz}}, \bibinfo {author}
  {\bibfnamefont {V.}~\bibnamefont {Bagnato}}, \bibinfo {author} {\bibfnamefont
  {R.}~\bibnamefont {Napolitano}}, \bibinfo {author} {\bibfnamefont
  {J.}~\bibnamefont {Weiner}}, \ and\ \bibinfo {author} {\bibfnamefont {P.~S.}\
  \bibnamefont {Julienne}},\ }\href@noop {} {\bibfield  {journal} {\bibinfo
  {journal} {Phys. Rev. Lett.}\ }\textbf {\bibinfo {volume} {76}},\ \bibinfo
  {pages} {2033} (\bibinfo {year} {1996})}\BibitemShut {NoStop}%
\bibitem [{\citenamefont {Suominen}\ \emph
  {et~al.}(1996{\natexlab{a}})\citenamefont {Suominen}, \citenamefont
  {Burnett}, \citenamefont {Julienne}, \citenamefont {Walhout}, \citenamefont
  {Sterr}, \citenamefont {Orzel}, \citenamefont {Hoogerland},\ and\
  \citenamefont {Rolston}}]{suominen1996a}%
  \BibitemOpen
  \bibfield  {author} {\bibinfo {author} {\bibfnamefont {K.-A.}\ \bibnamefont
  {Suominen}}, \bibinfo {author} {\bibfnamefont {K.}~\bibnamefont {Burnett}},
  \bibinfo {author} {\bibfnamefont {P.~S.}\ \bibnamefont {Julienne}}, \bibinfo
  {author} {\bibfnamefont {M.}~\bibnamefont {Walhout}}, \bibinfo {author}
  {\bibfnamefont {U.}~\bibnamefont {Sterr}}, \bibinfo {author} {\bibfnamefont
  {C.}~\bibnamefont {Orzel}}, \bibinfo {author} {\bibfnamefont
  {M.}~\bibnamefont {Hoogerland}}, \ and\ \bibinfo {author} {\bibfnamefont
  {S.~L.}\ \bibnamefont {Rolston}},\ }\href@noop {} {\bibfield  {journal}
  {\bibinfo  {journal} {Phys. Rev. A}\ }\textbf {\bibinfo {volume} {53}},\
  \bibinfo {pages} {1678} (\bibinfo {year} {1996}{\natexlab{a}})}\BibitemShut
  {NoStop}%
\bibitem [{\citenamefont {Weiner}\ \emph {et~al.}(1999)\citenamefont {Weiner},
  \citenamefont {Bagnato}, \citenamefont {Zilio},\ and\ \citenamefont
  {Julienne}}]{weiner1999}%
  \BibitemOpen
  \bibfield  {author} {\bibinfo {author} {\bibfnamefont {J.}~\bibnamefont
  {Weiner}}, \bibinfo {author} {\bibfnamefont {V.~S.}\ \bibnamefont {Bagnato}},
  \bibinfo {author} {\bibfnamefont {S.~C.}\ \bibnamefont {Zilio}}, \ and\
  \bibinfo {author} {\bibfnamefont {P.~S.}\ \bibnamefont {Julienne}},\
  }\href@noop {} {\bibfield  {journal} {\bibinfo  {journal} {Rev. Mod. Phys.}\
  }\textbf {\bibinfo {volume} {71}},\ \bibinfo {pages} {1} (\bibinfo {year}
  {1999})}\BibitemShut {NoStop}%
\bibitem [{\citenamefont {Suominen}\ \emph {et~al.}(1995)\citenamefont
  {Suominen}, \citenamefont {Holland}, \citenamefont {Burnett},\ and\
  \citenamefont {Julienne}}]{suominen1995}%
  \BibitemOpen
  \bibfield  {author} {\bibinfo {author} {\bibfnamefont {K.-A.}\ \bibnamefont
  {Suominen}}, \bibinfo {author} {\bibfnamefont {M.~J.}\ \bibnamefont
  {Holland}}, \bibinfo {author} {\bibfnamefont {K.}~\bibnamefont {Burnett}}, \
  and\ \bibinfo {author} {\bibfnamefont {P.}~\bibnamefont {Julienne}},\
  }\href@noop {} {\bibfield  {journal} {\bibinfo  {journal} {Phys. Rev. A}\
  }\textbf {\bibinfo {volume} {51}},\ \bibinfo {pages} {1446} (\bibinfo {year}
  {1995})}\BibitemShut {NoStop}%
\bibitem [{\citenamefont {Napolitano}\ \emph {et~al.}(1997)\citenamefont
  {Napolitano}, \citenamefont {Weiner},\ and\ \citenamefont
  {Julienne}}]{napolitano1997}%
  \BibitemOpen
  \bibfield  {author} {\bibinfo {author} {\bibfnamefont {R.}~\bibnamefont
  {Napolitano}}, \bibinfo {author} {\bibfnamefont {J.}~\bibnamefont {Weiner}},
  \ and\ \bibinfo {author} {\bibfnamefont {P.~S.}\ \bibnamefont {Julienne}},\
  }\href@noop {} {\bibfield  {journal} {\bibinfo  {journal} {Phys. Rev. A}\
  }\textbf {\bibinfo {volume} {55}},\ \bibinfo {pages} {1191} (\bibinfo {year}
  {1997})}\BibitemShut {NoStop}%
\bibitem [{\citenamefont {Walhout}\ \emph {et~al.}(1995)\citenamefont
  {Walhout}, \citenamefont {Sterr}, \citenamefont {Orzel}, \citenamefont
  {Hoogerland},\ and\ \citenamefont {Rolston}}]{walhout1995}%
  \BibitemOpen
  \bibfield  {author} {\bibinfo {author} {\bibfnamefont {M.}~\bibnamefont
  {Walhout}}, \bibinfo {author} {\bibfnamefont {U.}~\bibnamefont {Sterr}},
  \bibinfo {author} {\bibfnamefont {C.}~\bibnamefont {Orzel}}, \bibinfo
  {author} {\bibfnamefont {M.}~\bibnamefont {Hoogerland}}, \ and\ \bibinfo
  {author} {\bibfnamefont {S.}~\bibnamefont {Rolston}},\ }\href@noop {}
  {\bibfield  {journal} {\bibinfo  {journal} {Phys. Rev. Lett.}\ }\textbf
  {\bibinfo {volume} {74}},\ \bibinfo {pages} {506} (\bibinfo {year}
  {1995})}\BibitemShut {NoStop}%
\bibitem [{\citenamefont {Marcassa}\ \emph {et~al.}(1995)\citenamefont
  {Marcassa}, \citenamefont {Horowicz}, \citenamefont {Zilio}, \citenamefont
  {Bagnato},\ and\ \citenamefont {Weiner}}]{marcassa1995}%
  \BibitemOpen
  \bibfield  {author} {\bibinfo {author} {\bibfnamefont {L.}~\bibnamefont
  {Marcassa}}, \bibinfo {author} {\bibfnamefont {R.}~\bibnamefont {Horowicz}},
  \bibinfo {author} {\bibfnamefont {S.}~\bibnamefont {Zilio}}, \bibinfo
  {author} {\bibfnamefont {V.}~\bibnamefont {Bagnato}}, \ and\ \bibinfo
  {author} {\bibfnamefont {J.}~\bibnamefont {Weiner}},\ }\href@noop {}
  {\bibfield  {journal} {\bibinfo  {journal} {Phys. Rev. A}\ }\textbf {\bibinfo
  {volume} {52}},\ \bibinfo {pages} {R913} (\bibinfo {year}
  {1995})}\BibitemShut {NoStop}%
\bibitem [{\citenamefont {Orb\'an}\ \emph {et~al.}(2015)\citenamefont
  {Orb\'an}, \citenamefont {Vexiau}, \citenamefont {Krieglsteiner},
  \citenamefont {N\"agerl}, \citenamefont {Dulieu}, \citenamefont
  {Crubellier},\ and\ \citenamefont {Bouloufa-Maafa}}]{orban2015}%
  \BibitemOpen
  \bibfield  {author} {\bibinfo {author} {\bibfnamefont {A.}~\bibnamefont
  {Orb\'an}}, \bibinfo {author} {\bibfnamefont {R.}~\bibnamefont {Vexiau}},
  \bibinfo {author} {\bibfnamefont {O.}~\bibnamefont {Krieglsteiner}}, \bibinfo
  {author} {\bibfnamefont {H.-C.}\ \bibnamefont {N\"agerl}}, \bibinfo {author}
  {\bibfnamefont {O.}~\bibnamefont {Dulieu}}, \bibinfo {author} {\bibfnamefont
  {A.}~\bibnamefont {Crubellier}}, \ and\ \bibinfo {author} {\bibfnamefont
  {N.}~\bibnamefont {Bouloufa-Maafa}},\ }\href@noop {} {\bibfield  {journal}
  {\bibinfo  {journal} {Phys. Rev. A}\ }\textbf {\bibinfo {volume} {92}},\
  \bibinfo {pages} {032510} (\bibinfo {year} {2015})}\BibitemShut {NoStop}%
\bibitem [{\citenamefont {Borsalino}\ \emph {et~al.}(2016)\citenamefont
  {Borsalino}, \citenamefont {Vexiau}, \citenamefont {Aymar}, \citenamefont
  {Luc-Koenig}, \citenamefont {Dulieu},\ and\ \citenamefont
  {Bouloufa-Maafa}}]{borsalino2016}%
  \BibitemOpen
  \bibfield  {author} {\bibinfo {author} {\bibfnamefont {D.}~\bibnamefont
  {Borsalino}}, \bibinfo {author} {\bibfnamefont {R.}~\bibnamefont {Vexiau}},
  \bibinfo {author} {\bibfnamefont {M.}~\bibnamefont {Aymar}}, \bibinfo
  {author} {\bibfnamefont {E.}~\bibnamefont {Luc-Koenig}}, \bibinfo {author}
  {\bibfnamefont {O.}~\bibnamefont {Dulieu}}, \ and\ \bibinfo {author}
  {\bibfnamefont {N.}~\bibnamefont {Bouloufa-Maafa}},\ }\href@noop {}
  {\bibfield  {journal} {\bibinfo  {journal} {J. phys. B}\ }\textbf {\bibinfo
  {volume} {49}},\ \bibinfo {pages} {055301} (\bibinfo {year}
  {2016})}\BibitemShut {NoStop}%
\bibitem [{\citenamefont {Orb{\'{a}}n}\ \emph {et~al.}(2019)\citenamefont
  {Orb{\'{a}}n}, \citenamefont {Xie}, \citenamefont {Vexiau}, \citenamefont
  {Dulieu},\ and\ \citenamefont {Bouloufa-Maafa}}]{orban2019}%
  \BibitemOpen
  \bibfield  {author} {\bibinfo {author} {\bibfnamefont {A.}~\bibnamefont
  {Orb{\'{a}}n}}, \bibinfo {author} {\bibfnamefont {T.}~\bibnamefont {Xie}},
  \bibinfo {author} {\bibfnamefont {R.}~\bibnamefont {Vexiau}}, \bibinfo
  {author} {\bibfnamefont {O.}~\bibnamefont {Dulieu}}, \ and\ \bibinfo {author}
  {\bibfnamefont {N.}~\bibnamefont {Bouloufa-Maafa}},\ }\href@noop {}
  {\bibfield  {journal} {\bibinfo  {journal} {Journal of Physics B: Atomic,
  Molecular and Optical Physics}\ }\textbf {\bibinfo {volume} {52}},\ \bibinfo
  {pages} {135101} (\bibinfo {year} {2019})}\BibitemShut {NoStop}%
\bibitem [{\citenamefont {Gr\"obner}\ \emph {et~al.}(2017)\citenamefont
  {Gr\"obner}, \citenamefont {Weinmann}, \citenamefont {Kirilov}, \citenamefont
  {N\"agerl}, \citenamefont {Julienne}, \citenamefont {Le~Sueur},\ and\
  \citenamefont {Hutson}}]{groebner2017}%
  \BibitemOpen
  \bibfield  {author} {\bibinfo {author} {\bibfnamefont {M.}~\bibnamefont
  {Gr\"obner}}, \bibinfo {author} {\bibfnamefont {P.}~\bibnamefont {Weinmann}},
  \bibinfo {author} {\bibfnamefont {E.}~\bibnamefont {Kirilov}}, \bibinfo
  {author} {\bibfnamefont {H.-C.}\ \bibnamefont {N\"agerl}}, \bibinfo {author}
  {\bibfnamefont {P.~S.}\ \bibnamefont {Julienne}}, \bibinfo {author}
  {\bibfnamefont {C.~R.}\ \bibnamefont {Le~Sueur}}, \ and\ \bibinfo {author}
  {\bibfnamefont {J.~M.}\ \bibnamefont {Hutson}},\ }\href@noop {} {\bibfield
  {journal} {\bibinfo  {journal} {Phys. Rev. A}\ }\textbf {\bibinfo {volume}
  {95}},\ \bibinfo {pages} {022715} (\bibinfo {year} {2017})}\BibitemShut
  {NoStop}%
\bibitem [{\citenamefont {Marinescu}\ and\ \citenamefont
  {Sadeghpour}(1999)}]{marinescu1999}%
  \BibitemOpen
  \bibfield  {author} {\bibinfo {author} {\bibfnamefont {M.}~\bibnamefont
  {Marinescu}}\ and\ \bibinfo {author} {\bibfnamefont {H.~R.}\ \bibnamefont
  {Sadeghpour}},\ }\href@noop {} {\bibfield  {journal} {\bibinfo  {journal}
  {Phys. Rev. A}\ }\textbf {\bibinfo {volume} {59}},\ \bibinfo {pages} {390}
  (\bibinfo {year} {1999})}\BibitemShut {NoStop}%
\bibitem [{\citenamefont {Ferber}\ \emph {et~al.}(2013)\citenamefont {Ferber},
  \citenamefont {Nikolayeva}, \citenamefont {Tamanis}, \citenamefont
  {Kn\"ockel},\ and\ \citenamefont {Tiemann}}]{ferber2013}%
  \BibitemOpen
  \bibfield  {author} {\bibinfo {author} {\bibfnamefont {R.}~\bibnamefont
  {Ferber}}, \bibinfo {author} {\bibfnamefont {O.}~\bibnamefont {Nikolayeva}},
  \bibinfo {author} {\bibfnamefont {M.}~\bibnamefont {Tamanis}}, \bibinfo
  {author} {\bibfnamefont {H.}~\bibnamefont {Kn\"ockel}}, \ and\ \bibinfo
  {author} {\bibfnamefont {E.}~\bibnamefont {Tiemann}},\ }\href@noop {}
  {\bibfield  {journal} {\bibinfo  {journal} {Phys. Rev. A}\ }\textbf {\bibinfo
  {volume} {88}},\ \bibinfo {pages} {012516} (\bibinfo {year}
  {2013})}\BibitemShut {NoStop}%
\bibitem [{\citenamefont {Nicholson}\ \emph {et~al.}(2015)\citenamefont
  {Nicholson}, \citenamefont {Blatt}, \citenamefont {Bloom}, \citenamefont
  {Williams}, \citenamefont {Thomsen}, \citenamefont {Ye},\ and\ \citenamefont
  {Julienne}}]{nicholson2015}%
  \BibitemOpen
  \bibfield  {author} {\bibinfo {author} {\bibfnamefont {T.~L.}\ \bibnamefont
  {Nicholson}}, \bibinfo {author} {\bibfnamefont {S.}~\bibnamefont {Blatt}},
  \bibinfo {author} {\bibfnamefont {B.~J.}\ \bibnamefont {Bloom}}, \bibinfo
  {author} {\bibfnamefont {J.~R.}\ \bibnamefont {Williams}}, \bibinfo {author}
  {\bibfnamefont {J.~W.}\ \bibnamefont {Thomsen}}, \bibinfo {author}
  {\bibfnamefont {J.}~\bibnamefont {Ye}}, \ and\ \bibinfo {author}
  {\bibfnamefont {P.~S.}\ \bibnamefont {Julienne}},\ }\href@noop {} {\bibfield
  {journal} {\bibinfo  {journal} {Phys. Rev. A}\ }\textbf {\bibinfo {volume}
  {92}},\ \bibinfo {pages} {022709} (\bibinfo {year} {2015})}\BibitemShut
  {NoStop}%
\bibitem [{\citenamefont {Julienne}\ and\ \citenamefont
  {Mies}(1984)}]{julienne1984}%
  \BibitemOpen
  \bibfield  {author} {\bibinfo {author} {\bibfnamefont {P.~S.}\ \bibnamefont
  {Julienne}}\ and\ \bibinfo {author} {\bibfnamefont {F.~H.}\ \bibnamefont
  {Mies}},\ }\href@noop {} {\bibfield  {journal} {\bibinfo  {journal} {Phys.
  Rev. A}\ }\textbf {\bibinfo {volume} {30}},\ \bibinfo {pages} {831} (\bibinfo
  {year} {1984})}\BibitemShut {NoStop}%
\bibitem [{\citenamefont {Drozdova}(2012)}]{drozdova2012}%
  \BibitemOpen
  \bibfield  {author} {\bibinfo {author} {\bibfnamefont {A.}~\bibnamefont
  {Drozdova}},\ }\emph {\bibinfo {title} {Study of spin-orbit coupled
  electronic states of Rb$_2$, NaCs and NaK molecules: Laser spectroscopy and
  accurate coupled-channel deperturbation analysis}},\ \href@noop {} {Ph.D.
  thesis},\ \bibinfo {address} {Universit\'{e} Claude Bernard - Lyon I, France}
  (\bibinfo {year} {2012}),\ \bibinfo {note}
  {https://tel.archives-ouvertes.fr/tel-01127557}\BibitemShut {NoStop}%
\bibitem [{\citenamefont {Singer}\ \emph {et~al.}(1983)\citenamefont {Singer},
  \citenamefont {Freed},\ and\ \citenamefont {Band}}]{singer1983}%
  \BibitemOpen
  \bibfield  {author} {\bibinfo {author} {\bibfnamefont {S.~J.}\ \bibnamefont
  {Singer}}, \bibinfo {author} {\bibfnamefont {K.~F.}\ \bibnamefont {Freed}}, \
  and\ \bibinfo {author} {\bibfnamefont {Y.~B.}\ \bibnamefont {Band}},\
  }\href@noop {} {\bibfield  {journal} {\bibinfo  {journal} {J. Chem. Phys.}\
  }\textbf {\bibinfo {volume} {79}},\ \bibinfo {pages} {6060} (\bibinfo {year}
  {1983})}\BibitemShut {NoStop}%
\bibitem [{\citenamefont {Dubs}\ and\ \citenamefont
  {Julienne}(1991)}]{dubs1991}%
  \BibitemOpen
  \bibfield  {author} {\bibinfo {author} {\bibfnamefont {R.~L.}\ \bibnamefont
  {Dubs}}\ and\ \bibinfo {author} {\bibfnamefont {P.~S.}\ \bibnamefont
  {Julienne}},\ }\href@noop {} {\bibfield  {journal} {\bibinfo  {journal} {J.
  Chem. Phys.}\ }\textbf {\bibinfo {volume} {95}},\ \bibinfo {pages} {4177}
  (\bibinfo {year} {1991})}\BibitemShut {NoStop}%
\bibitem [{\citenamefont {Bergeman}\ \emph {et~al.}(2002)\citenamefont
  {Bergeman}, \citenamefont {Julienne}, \citenamefont {Williams}, \citenamefont
  {Tiesinga}, \citenamefont {Manaa}, \citenamefont {Wang}, \citenamefont
  {Gould},\ and\ \citenamefont {Stwalley}}]{bergeman2002}%
  \BibitemOpen
  \bibfield  {author} {\bibinfo {author} {\bibfnamefont {T.}~\bibnamefont
  {Bergeman}}, \bibinfo {author} {\bibfnamefont {P.~S.}\ \bibnamefont
  {Julienne}}, \bibinfo {author} {\bibfnamefont {C.~J.}\ \bibnamefont
  {Williams}}, \bibinfo {author} {\bibfnamefont {E.}~\bibnamefont {Tiesinga}},
  \bibinfo {author} {\bibfnamefont {M.~R.}\ \bibnamefont {Manaa}}, \bibinfo
  {author} {\bibfnamefont {H.}~\bibnamefont {Wang}}, \bibinfo {author}
  {\bibfnamefont {P.~L.}\ \bibnamefont {Gould}}, \ and\ \bibinfo {author}
  {\bibfnamefont {W.~C.}\ \bibnamefont {Stwalley}},\ }\href@noop {} {\bibfield
  {journal} {\bibinfo  {journal} {J. Chem. Phys.}\ }\textbf {\bibinfo {volume}
  {117}},\ \bibinfo {pages} {7491} (\bibinfo {year} {2002})}\BibitemShut
  {NoStop}%
\bibitem [{\citenamefont {Napolitano}(1998)}]{napolitano1998}%
  \BibitemOpen
  \bibfield  {author} {\bibinfo {author} {\bibfnamefont {R.}~\bibnamefont
  {Napolitano}},\ }\href@noop {} {\bibfield  {journal} {\bibinfo  {journal}
  {Phys. Rev. A}\ }\textbf {\bibinfo {volume} {57}},\ \bibinfo {pages} {1164}
  (\bibinfo {year} {1998})}\BibitemShut {NoStop}%
\bibitem [{\citenamefont {Julienne}\ \emph {et~al.}(1994)\citenamefont
  {Julienne}, \citenamefont {Suominen},\ and\ \citenamefont
  {Band}}]{julienne1994}%
  \BibitemOpen
  \bibfield  {author} {\bibinfo {author} {\bibfnamefont {P.~S.}\ \bibnamefont
  {Julienne}}, \bibinfo {author} {\bibfnamefont {K.-A.}\ \bibnamefont
  {Suominen}}, \ and\ \bibinfo {author} {\bibfnamefont {Y.}~\bibnamefont
  {Band}},\ }\href@noop {} {\bibfield  {journal} {\bibinfo  {journal} {Phys.
  Rev. A}\ }\textbf {\bibinfo {volume} {49}},\ \bibinfo {pages} {3890}
  (\bibinfo {year} {1994})}\BibitemShut {NoStop}%
\bibitem [{\citenamefont {Johnson}(1973)}]{johnson1973}%
  \BibitemOpen
  \bibfield  {author} {\bibinfo {author} {\bibfnamefont {B.}~\bibnamefont
  {Johnson}},\ }\href@noop {} {\bibfield  {journal} {\bibinfo  {journal} {J.
  Comp. Phys.}\ }\textbf {\bibinfo {volume} {13}},\ \bibinfo {pages} {445}
  (\bibinfo {year} {1973})}\BibitemShut {NoStop}%
\bibitem [{\citenamefont {Tuvi}\ and\ \citenamefont {Band}(1993)}]{tuvi1993}%
  \BibitemOpen
  \bibfield  {author} {\bibinfo {author} {\bibfnamefont {I.}~\bibnamefont
  {Tuvi}}\ and\ \bibinfo {author} {\bibfnamefont {Y.~B.}\ \bibnamefont
  {Band}},\ }\href@noop {} {\bibfield  {journal} {\bibinfo  {journal} {J. Chem.
  Phys.}\ }\textbf {\bibinfo {volume} {99}},\ \bibinfo {pages} {9697} (\bibinfo
  {year} {1993})}\BibitemShut {NoStop}%
\bibitem [{\citenamefont {Aymar}\ and\ \citenamefont
  {Dulieu}(2005)}]{aymar2005}%
  \BibitemOpen
  \bibfield  {author} {\bibinfo {author} {\bibfnamefont {M.}~\bibnamefont
  {Aymar}}\ and\ \bibinfo {author} {\bibfnamefont {O.}~\bibnamefont {Dulieu}},\
  }\href@noop {} {\bibfield  {journal} {\bibinfo  {journal} {J. Chem. Phys.}\
  }\textbf {\bibinfo {volume} {122}},\ \bibinfo {pages} {204302} (\bibinfo
  {year} {2005})}\BibitemShut {NoStop}%
\bibitem [{\citenamefont {Aymar}\ \emph {et~al.}(2006)\citenamefont {Aymar},
  \citenamefont {Dulieu},\ and\ \citenamefont {Spiegelman}}]{aymar2006}%
  \BibitemOpen
  \bibfield  {author} {\bibinfo {author} {\bibfnamefont {M.}~\bibnamefont
  {Aymar}}, \bibinfo {author} {\bibfnamefont {O.}~\bibnamefont {Dulieu}}, \
  and\ \bibinfo {author} {\bibfnamefont {F.}~\bibnamefont {Spiegelman}},\
  }\href@noop {} {\bibfield  {journal} {\bibinfo  {journal} {J. Phys. B: At.
  Mol. Opt. Phys.}\ }\textbf {\bibinfo {volume} {39}},\ \bibinfo {pages} {S905}
  (\bibinfo {year} {2006})}\BibitemShut {NoStop}%
\bibitem [{\citenamefont {Kramida}\ \emph {et~al.}(2018)\citenamefont
  {Kramida}, \citenamefont {{Yu.~Ralchenko}}, \citenamefont {Reader},\ and\
  \citenamefont {{and NIST ASD Team}}}]{NIST_ASD}%
  \BibitemOpen
  \bibfield  {author} {\bibinfo {author} {\bibfnamefont {A.}~\bibnamefont
  {Kramida}}, \bibinfo {author} {\bibnamefont {{Yu.~Ralchenko}}}, \bibinfo
  {author} {\bibfnamefont {J.}~\bibnamefont {Reader}}, \ and\ \bibinfo {author}
  {\bibnamefont {{and NIST ASD Team}}},\ }\href@noop {} {}\bibinfo
  {howpublished} {{NIST Atomic Spectra Database (ver. 5.6.1), [Online].
  Available: {\tt{https://physics.nist.gov/asd}} [2019, March 28]. National
  Institute of Standards and Technology, Gaithersburg, MD.}} (\bibinfo {year}
  {2018})\BibitemShut {NoStop}%
\bibitem [{\citenamefont {Wang}\ \emph {et~al.}(1996)\citenamefont {Wang},
  \citenamefont {Gould},\ and\ \citenamefont {Stwalley}}]{wang1996}%
  \BibitemOpen
  \bibfield  {author} {\bibinfo {author} {\bibfnamefont {H.}~\bibnamefont
  {Wang}}, \bibinfo {author} {\bibfnamefont {P.~L.}\ \bibnamefont {Gould}}, \
  and\ \bibinfo {author} {\bibfnamefont {W.}~\bibnamefont {Stwalley}},\
  }\href@noop {} {\bibfield  {journal} {\bibinfo  {journal} {Phys. Rev. A}\
  }\textbf {\bibinfo {volume} {53}},\ \bibinfo {pages} {R1216} (\bibinfo {year}
  {1996})}\BibitemShut {NoStop}%
\bibitem [{\citenamefont {Suominen}\ \emph
  {et~al.}(1996{\natexlab{b}})\citenamefont {Suominen}, \citenamefont
  {Burnett},\ and\ \citenamefont {Julienne}}]{suominen1996b}%
  \BibitemOpen
  \bibfield  {author} {\bibinfo {author} {\bibfnamefont {K.~A.}\ \bibnamefont
  {Suominen}}, \bibinfo {author} {\bibfnamefont {K.}~\bibnamefont {Burnett}}, \
  and\ \bibinfo {author} {\bibfnamefont {P.~S.}\ \bibnamefont {Julienne}},\
  }\href@noop {} {\bibfield  {journal} {\bibinfo  {journal} {Phys. Rev. A}\
  }\textbf {\bibinfo {volume} {53}},\ \bibinfo {pages} {R1220} (\bibinfo {year}
  {1996}{\natexlab{b}})}\BibitemShut {NoStop}%
\bibitem [{\citenamefont {Santos}\ \emph {et~al.}(1999)\citenamefont {Santos},
  \citenamefont {Nussenzveig}, \citenamefont {Antunes}, \citenamefont
  {Cardona},\ and\ \citenamefont {Bagnato}}]{santos1999}%
  \BibitemOpen
  \bibfield  {author} {\bibinfo {author} {\bibfnamefont {M.~S.}\ \bibnamefont
  {Santos}}, \bibinfo {author} {\bibfnamefont {P.}~\bibnamefont {Nussenzveig}},
  \bibinfo {author} {\bibfnamefont {A.}~\bibnamefont {Antunes}}, \bibinfo
  {author} {\bibfnamefont {P.~S.~P.}\ \bibnamefont {Cardona}}, \ and\ \bibinfo
  {author} {\bibfnamefont {V.~S.}\ \bibnamefont {Bagnato}},\ }\href@noop {}
  {\bibfield  {journal} {\bibinfo  {journal} {Phys. Rev. A}\ }\textbf {\bibinfo
  {volume} {60}},\ \bibinfo {pages} {3892} (\bibinfo {year}
  {1999})}\BibitemShut {NoStop}%
\bibitem [{\citenamefont {Young}\ \emph {et~al.}(2000)\citenamefont {Young},
  \citenamefont {Ejnisman}, \citenamefont {Shaffer},\ and\ \citenamefont
  {Bigelow}}]{young2000}%
  \BibitemOpen
  \bibfield  {author} {\bibinfo {author} {\bibfnamefont {Y.~E.}\ \bibnamefont
  {Young}}, \bibinfo {author} {\bibfnamefont {R.}~\bibnamefont {Ejnisman}},
  \bibinfo {author} {\bibfnamefont {J.~P.}\ \bibnamefont {Shaffer}}, \ and\
  \bibinfo {author} {\bibfnamefont {N.~P.}\ \bibnamefont {Bigelow}},\
  }\href@noop {} {\bibfield  {journal} {\bibinfo  {journal} {Phys. Rev. A}\
  }\textbf {\bibinfo {volume} {62}},\ \bibinfo {pages} {055403} (\bibinfo
  {year} {2000})}\BibitemShut {NoStop}%
\bibitem [{\citenamefont {Marcassa}\ \emph {et~al.}(2000)\citenamefont
  {Marcassa}, \citenamefont {Telles}, \citenamefont {Muniz},\ and\
  \citenamefont {Bagnato}}]{marcassa2000}%
  \BibitemOpen
  \bibfield  {author} {\bibinfo {author} {\bibfnamefont {L.~G.}\ \bibnamefont
  {Marcassa}}, \bibinfo {author} {\bibfnamefont {G.~D.}\ \bibnamefont
  {Telles}}, \bibinfo {author} {\bibfnamefont {S.~R.}\ \bibnamefont {Muniz}}, \
  and\ \bibinfo {author} {\bibfnamefont {V.~S.}\ \bibnamefont {Bagnato}},\
  }\href@noop {} {\bibfield  {journal} {\bibinfo  {journal} {Phys. Rev. A}\
  }\textbf {\bibinfo {volume} {63}},\ \bibinfo {pages} {013413} (\bibinfo
  {year} {2000})}\BibitemShut {NoStop}%
\bibitem [{\citenamefont {Mudrich}\ \emph {et~al.}(2004)\citenamefont
  {Mudrich}, \citenamefont {Kraft}, \citenamefont {Lange}, \citenamefont
  {Mosk}, \citenamefont {Weidem\"uller},\ and\ \citenamefont
  {Tiesinga}}]{mudrich2004b}%
  \BibitemOpen
  \bibfield  {author} {\bibinfo {author} {\bibfnamefont {M.}~\bibnamefont
  {Mudrich}}, \bibinfo {author} {\bibfnamefont {S.}~\bibnamefont {Kraft}},
  \bibinfo {author} {\bibfnamefont {J.}~\bibnamefont {Lange}}, \bibinfo
  {author} {\bibfnamefont {A.}~\bibnamefont {Mosk}}, \bibinfo {author}
  {\bibfnamefont {M.}~\bibnamefont {Weidem\"uller}}, \ and\ \bibinfo {author}
  {\bibfnamefont {E.}~\bibnamefont {Tiesinga}},\ }\href@noop {} {\bibfield
  {journal} {\bibinfo  {journal} {Phys. Rev. A}\ }\textbf {\bibinfo {volume}
  {70}},\ \bibinfo {pages} {062712} (\bibinfo {year} {2004})}\BibitemShut
  {NoStop}%
\bibitem [{\citenamefont {Wang}\ \emph {et~al.}(2020)\citenamefont {Wang},
  \citenamefont {Wang}, \citenamefont {Liu}, \citenamefont {Qi}, \citenamefont
  {Yao}, \citenamefont {Chen},\ and\ \citenamefont {Pan}}]{wang2020}%
  \BibitemOpen
  \bibfield  {author} {\bibinfo {author} {\bibfnamefont {X.-Q.}\ \bibnamefont
  {Wang}}, \bibinfo {author} {\bibfnamefont {Y.-X.}\ \bibnamefont {Wang}},
  \bibinfo {author} {\bibfnamefont {X.-P.}\ \bibnamefont {Liu}}, \bibinfo
  {author} {\bibfnamefont {R.}~\bibnamefont {Qi}}, \bibinfo {author}
  {\bibfnamefont {X.-C.}\ \bibnamefont {Yao}}, \bibinfo {author} {\bibfnamefont
  {Y.-A.}\ \bibnamefont {Chen}}, \ and\ \bibinfo {author} {\bibfnamefont
  {J.-W.}\ \bibnamefont {Pan}},\ }\href@noop {} {\bibfield  {journal} {\bibinfo
   {journal} {Phys. Rev. A}\ }\textbf {\bibinfo {volume} {101}},\ \bibinfo
  {pages} {041601} (\bibinfo {year} {2020})}\BibitemShut {NoStop}%
\bibitem [{\citenamefont {Beuc}\ \emph {et~al.}(2006)\citenamefont {Beuc},
  \citenamefont {Movre}, \citenamefont {Ban}, \citenamefont {Pichler},
  \citenamefont {Aymar}, \citenamefont {Dulieu},\ and\ \citenamefont
  {Ernst}}]{beuc2006}%
  \BibitemOpen
  \bibfield  {author} {\bibinfo {author} {\bibfnamefont {R.}~\bibnamefont
  {Beuc}}, \bibinfo {author} {\bibfnamefont {M.}~\bibnamefont {Movre}},
  \bibinfo {author} {\bibfnamefont {T.}~\bibnamefont {Ban}}, \bibinfo {author}
  {\bibfnamefont {C.}~\bibnamefont {Pichler}}, \bibinfo {author} {\bibfnamefont
  {M.}~\bibnamefont {Aymar}}, \bibinfo {author} {\bibfnamefont
  {O.}~\bibnamefont {Dulieu}}, \ and\ \bibinfo {author} {\bibfnamefont {W.~E.}\
  \bibnamefont {Ernst}},\ }\href@noop {} {\bibfield  {journal} {\bibinfo
  {journal} {J. Phys. B: At. Mol. Opt. Phys.}\ }\textbf {\bibinfo {volume}
  {39}},\ \bibinfo {pages} {S1191} (\bibinfo {year} {2006})}\BibitemShut
  {NoStop}%
\bibitem [{\citenamefont {Raki\'{c}}\ \emph {et~al.}(2016)\citenamefont
  {Raki\'{c}}, \citenamefont {Beuc}, \citenamefont {Bouloufa-Maafa},
  \citenamefont {Dulieu}, \citenamefont {Vexiau}, \citenamefont {Pichler},\
  and\ \citenamefont {Skenderovi\'{c}}}]{beuc2016}%
  \BibitemOpen
  \bibfield  {author} {\bibinfo {author} {\bibfnamefont {M.}~\bibnamefont
  {Raki\'{c}}}, \bibinfo {author} {\bibfnamefont {R.}~\bibnamefont {Beuc}},
  \bibinfo {author} {\bibfnamefont {N.}~\bibnamefont {Bouloufa-Maafa}},
  \bibinfo {author} {\bibfnamefont {O.}~\bibnamefont {Dulieu}}, \bibinfo
  {author} {\bibfnamefont {R.}~\bibnamefont {Vexiau}}, \bibinfo {author}
  {\bibfnamefont {G.}~\bibnamefont {Pichler}}, \ and\ \bibinfo {author}
  {\bibfnamefont {H.}~\bibnamefont {Skenderovi\'{c}}},\ }\href@noop {}
  {\bibfield  {journal} {\bibinfo  {journal} {J. Chem. Phys}\ }\textbf
  {\bibinfo {volume} {144}},\ \bibinfo {pages} {204310} (\bibinfo {year}
  {2016})}\BibitemShut {NoStop}%
\bibitem [{\citenamefont {Burchianti}\ \emph {et~al.}(2018)\citenamefont
  {Burchianti}, \citenamefont {D'Errico}, \citenamefont {Rosi}, \citenamefont
  {Simoni}, \citenamefont {Modugno}, \citenamefont {Fort},\ and\ \citenamefont
  {Minardi}}]{burchianti2018}%
  \BibitemOpen
  \bibfield  {author} {\bibinfo {author} {\bibfnamefont {A.}~\bibnamefont
  {Burchianti}}, \bibinfo {author} {\bibfnamefont {C.}~\bibnamefont
  {D'Errico}}, \bibinfo {author} {\bibfnamefont {S.}~\bibnamefont {Rosi}},
  \bibinfo {author} {\bibfnamefont {A.}~\bibnamefont {Simoni}}, \bibinfo
  {author} {\bibfnamefont {M.}~\bibnamefont {Modugno}}, \bibinfo {author}
  {\bibfnamefont {C.}~\bibnamefont {Fort}}, \ and\ \bibinfo {author}
  {\bibfnamefont {F.}~\bibnamefont {Minardi}},\ }\href@noop {} {\bibfield
  {journal} {\bibinfo  {journal} {Phys. Rev. A}\ }\textbf {\bibinfo {volume}
  {98}},\ \bibinfo {pages} {063616} (\bibinfo {year} {2018})}\BibitemShut
  {NoStop}%
\end{thebibliography}

%

\end{document}